\documentclass[aps,pra,reprint,showpacs,superscriptaddress]{revtex4-1}

\usepackage{graphicx}
\usepackage{amsmath}
\usepackage{amssymb}
\usepackage[overlay,absolute]{textpos}
\usepackage[dvipdfm,bookmarksnumbered=true,pdfborder={0 0 0},pdfpagemode=UseNone,pdfstartview=FitH]{hyperref}

\begin{document}

\begin{textblock*}{\textwidth}[0,0](19mm,7mm)
\center\footnotesize\noindent
To appear in Phys.\ Rev.\ A: 
\href{https://journals.aps.org/pra/}{https://journals.aps.org/pra/}
\end{textblock*}

\title{Optical lattice implementation scheme of a bosonic topological model\\ with fermionic atoms}
\author{Anne E. B. Nielsen}
\affiliation{Max-Planck-Institut f{\"u}r Quantenoptik, Hans-Kopfermann-Str.\ 1, D-85748 Garching, Germany}
\author{Germ\'an Sierra}
\affiliation{Instituto de F\'isica Te\'orica, UAM-CSIC, Madrid, Spain}
\author{J. Ignacio Cirac}
\affiliation{Max-Planck-Institut f{\"u}r Quantenoptik, Hans-Kopfermann-Str.\ 1, D-85748 Garching, Germany}

\begin{abstract}
We present a scheme to implement a Fermi-Hubbard-like model in ultracold atoms in optical lattices and analyze the topological features of its ground state. In particular, we show that the ground state for appropriate parameters has a large overlap with a lattice version of the bosonic Laughlin state at filling factor one half. The scheme utilizes laser assisted and normal tunneling in a checkerboard optical lattice. The requirements on temperature, interactions, and hopping strengths are similar to those needed to observe the N\'eel antiferromagnetic ordering in the standard Fermi-Hubbard model in the Mott insulating regime.
\end{abstract}


\maketitle

\section{Introduction}\label{introduction}

Topological states have many interesting features with possible practical applications, and they are currently one of the main topics in strongly correlated many-body systems. The fractional quantum Hall (FQH) states of electrons in solids play a central role in this respect, because they are among the few cases, where topological states have been prepared experimentally, and because there is a quite detailed analytical understanding of the physics. After the discovery of the quantum Hall effect in solids \cite{tsui}, much work has been done to find similar behavior in other systems, in particular in lattices \cite{wen,schroeter,thomale,nielsen2,parameswaran,tu,maciejko}, where much less is currently known. The aim is to get a more detailed understanding of the nature of quantum Hall physics and to find alternative routes to realize it experimentally. Lattice systems are natural to investigate because they have long been used as toy models for understanding phenomena in condensed matter systems, and numerical computations are easier to accomplish on lattices. Another important motivation is the ongoing experimental progress in simulating quantum lattice models with ultracold atoms in optical lattices \cite{toolbox,blochrev,mazza}. Realizing FQH states in such systems would be very interesting, because the systems allow for a high degree of tunability, and with sophisticated techniques it is even possible to access the states at the single particle level \cite{weitenberg}.

A main strategy used so far to search for quantum-Hall-like states in lattices is to mimic characteristic features of the continuum setting, in which the quantum Hall effect was first observed, i.e., to find lattice replacements for the strong magnetic field, the quantized Hall conductivity, and the Landau levels \cite{roy,bergholtz}. A first step in this direction is to notice that the Aharonov-Bohm phase of charged particles moving in a magnetic field can be mimicked in lattice systems by introducing hopping terms in the Hamiltonian with complex hopping amplitudes that vary in space in such a way that a particle acquires a certain phase factor when it hops around some closed loop on the lattice \cite{haldane}. In the quasi continuum limit, in which the number of lattice sites is much larger than the number of flux lines and much larger than the number of atoms, such ideas are sufficient to achieve FQH-like behavior \cite{sorensen,hafezi,kapit}. The Hall conductivity in the continuum has turned out \cite{thouless,niu,kohmoto} to be closely related to a topological quantity called the (first) Chern number, and the Chern number can also be computed for lattice models \cite{hatsugai}. Haldane proposed a model \cite{haldane} with a nonzero Chern number and integer band filling that can be seen as a lattice version of the integer quantum Hall effect. The energy bands of this model are not flat like Landau levels, but this is not important as long as the bands are either completely filled or empty. To achieve FQH-like states, however, it is natural to expect that at least the partially filled band should be flat. Flattening can be achieved by fine tuning local hopping amplitudes \cite{tang,sun,neupert}, and theoretical studies for fractional filling predict that FQH-like states indeed appears if interactions are added \cite{neupert,sheng,wang,regnault}. Very recently proposals for how to implement such models experimentally have also appeared \cite{cooper,yao}.

In the present paper, we give a detailed description of a scheme \cite{nielsen3} to implement a lattice version of the bosonic Laughlin state at filling factor $\nu=1/2$ in ultracold fermionic atoms in optical lattices. We do this by showing that the state appears as the ground state of a Fermi-Hubbard-like model in the Mott insulating regime, which can be realized by using a combination of laser assisted \cite{jaksch,lin,aidelsburger,miyake} and normal tunneling in a checkerboard optical lattice. We also analyze the Fermi-Hubbard-like model and find that it is of a different type than the models described above, which suggests that FQH-like behavior can be obtained by other mechanisms than mimicking the continuum FQH setting.

The proposed setup requires eight laser beams for the trapping in the $xy$-plane and three additional standing wave laser fields to accomplish the hopping terms and the trapping in the $z$-direction. A particularly convenient feature of the scheme is that we do not need to implement interactions between atoms on different sites, since only on-site interactions are present. The requirements regarding temperatures, tunneling strengths, and interactions are the same as those needed to observe the N\'eel antiferromagnetic ordering in the normal Fermi-Hubbard model in the Mott insulating regime. More groups are already working on the latter, due to its expected relation to high $T_c$ superconductivity and to observe quantum magnetism \cite{lee,jordens,schneider,trotzky,blochNP,greif}. Our proposal can thus be implemented with present or planned technologies.

In Sec.~\ref{sec:model}, we introduce the Fermi-Hubbard-like model, show how it is related to a lattice version of the $\nu=1/2$ Laughlin state, and compute flatness and Chern number of the kinetic energy part of the model. The implementation scheme is described in Sec.~\ref{sec:implementation}, where we first give an overview of the ideas and then describe the implementation of the required optical lattice, the hopping terms, and the interaction terms in more detail. Section~\ref{sec:conclusion} concludes the paper, and the appendices provide further mathematical details.

\section{Model}\label{sec:model}

\subsection{The Fermi-Hubbard-like model}

\begin{figure}
\includegraphics[width=\columnwidth]{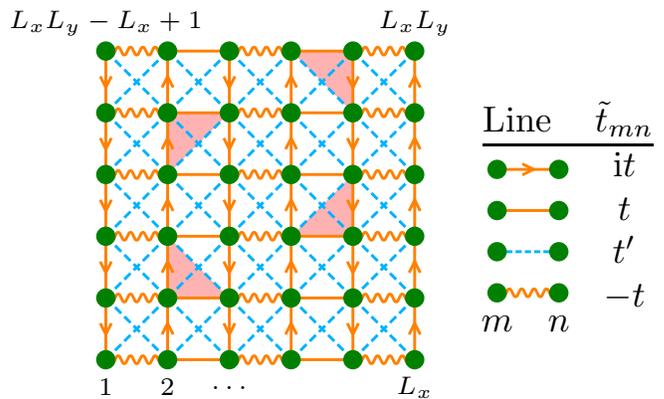}
\caption{(Color online) Amplitudes $\tilde{t}_{mn}$ of the hopping terms \eqref{Hkin} in the Fermi-Hubbard-like Hamiltonian \eqref{FH}. The green bullets are the lattice sites, and each line combining two sites represents that a fermion can hop between the sites with the amplitude given in the table on the right, where $t$ and $t'$ are real numbers. The rule for choosing the phases is that the product of the three $\tilde{t}$ factors appearing in the hopping terms that move a fermion in the counter clockwise direction around a triangle consisting of one horizontal line, one vertical line, and one diagonal line (i.e., any triangle that is a lattice translation of any of the four triangles marked in the figure) should always be $-i t^2t'$. Unless otherwise specified, we number the lattice sites rowwise starting from the lower left corner as shown.}\label{fig:Hkin}
\end{figure}

We consider fermions with spin on an $L_x\times L_y$ square lattice. We shall assume throughout that $L_x$ is even and that there are $N/2$ spin up fermions and $N/2$ spin down fermions, where $N=L_xL_y$ is the number of lattice sites. The Fermi-Hubbard-like Hamiltonian
\begin{equation}\label{FH}
H_{\rm FH}=\sum_{\sigma\in\{{\downarrow},{\uparrow}\}} H_{{\rm kin},\sigma} +H_{\rm int}
\end{equation}
that we consider consists of independent kinetic energy hopping terms
\begin{equation}\label{Hkin}
H_{{\rm kin},\sigma}=\sum_{\substack{{<n,m>}\\{<\!\!<n,m>\!\!>}}} (\tilde{t}_{mn}a_{n\sigma}^\dag a_{m\sigma}+\tilde{t}^*_{mn}a^\dag_{m\sigma}a_{n\sigma})
\end{equation}
for spin up and spin down and an on-site, repulsive interaction term
\begin{equation}\label{Hint}
H_{\rm int}=U\sum_{n=1}^N a_{n{\uparrow}}^\dag a_{n{\uparrow}}
a_{n{\downarrow}}^\dag a_{n{\downarrow}}.
\end{equation}
Here, $\tilde{t}_{mn}$ are the (complex) hopping amplitudes specified in Fig.~\ref{fig:Hkin}, $a_{n\sigma}$ is the annihilation operator of a fermion with spin $\sigma$ on site number $n$ (we number the sites rowwise starting from the lower left corner of the lattice as in Fig.~\ref{fig:Hkin}), the sum in \eqref{Hkin} is over all pairs of nearest and next-nearest neighbors on the lattice (using open boundary conditions), and $U$ is a real, positive constant.

Let us note that \eqref{FH} is $SU(2)$ invariant. This can be seen by writing
\begin{multline}
a_{n{\uparrow}}^\dag a_{n{\uparrow}}
a_{n{\downarrow}}^\dag a_{n{\downarrow}}=\\
\frac{1}{2}\left(\sum_\sigma a_{n\sigma}^\dag a_{n\sigma}
\sum_{\sigma'} a_{n\sigma'}^\dag a_{n\sigma'}
-\sum_\sigma a_{n\sigma}^\dag a_{n\sigma}\right).
\end{multline}
All terms in the Hamiltonian can thus be expressed in terms of $\sum_\sigma a^\dag_{n\sigma}a_{m\sigma}$. If the operators $a^\dag_{1\sigma}$, $a^\dag_{2\sigma}$, $\ldots$, $a^\dag_{N\sigma}$ are all transformed by the same unitary transformation acting on the index $\sigma$, then the action on $a^\dag_{n\sigma}$ and $a_{m\sigma}$ cancel each other, and this gives the $SU(2)$ invariance.

\subsection{Connection to the $\nu=1/2$ Laughlin state}

In the present paper, we are particularly interested in the limit of strong interactions, i.e.\ $|t|\ll U$ and $|t'|\ll U$, and we assume half filling of both spin up and spin down. In this case, it costs a lot of energy to put two fermions (with opposite spins) on the same site, and the low energy states are thus those with precisely one fermion on each lattice site. The low energy physics of the model is then given by an effective Hamiltonian $H_{\rm eff}$ acting on the low energy subspace, which can be derived by applying the Schrieffer-Wolff transformation (see Appendix \ref{sec:SW}). The effective model is a spin model because the basis states in the low energy subspace can be written as $|\sigma_1\sigma_2\ldots\sigma_N\rangle$, where $\sigma_n\in\{{\uparrow},{\downarrow}\}$ is the spin of the fermion at site $n$. We can also define spin operators $\vec{S}_n=(S_n^x,S_n^y,S_n^z)$ acting on the spin at site $n$ with standard spin commutation relations $[S_n^a,S_m^b]=i \delta_{nm}\sum_c\varepsilon_{abc}S_n^c$, where $\varepsilon_{abc}$ is the Levi-Civita symbol and $a,b,c\in\{x,y,z\}$.

\begin{figure*}
\includegraphics[width=0.32\textwidth]{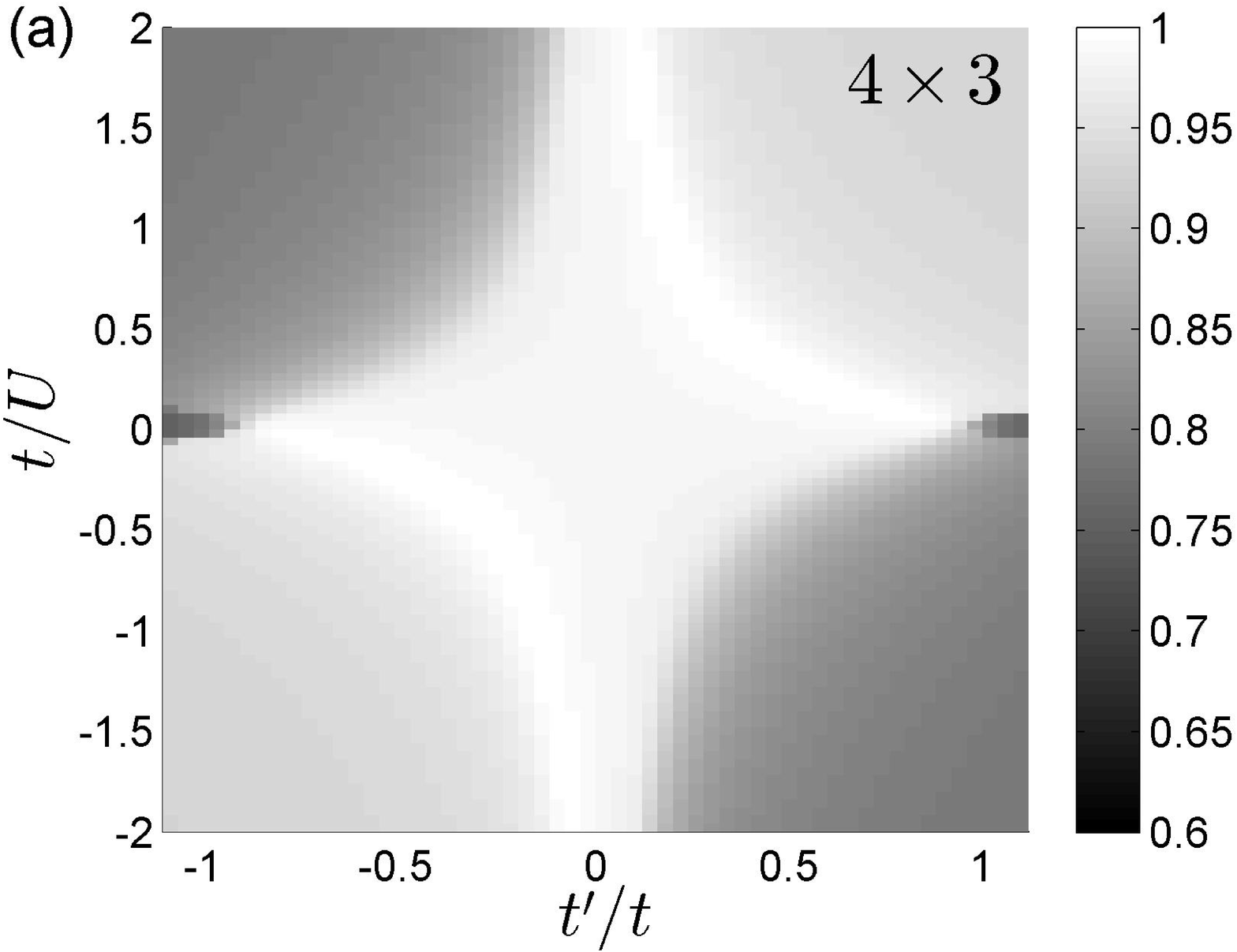}
\includegraphics[width=0.32\textwidth]{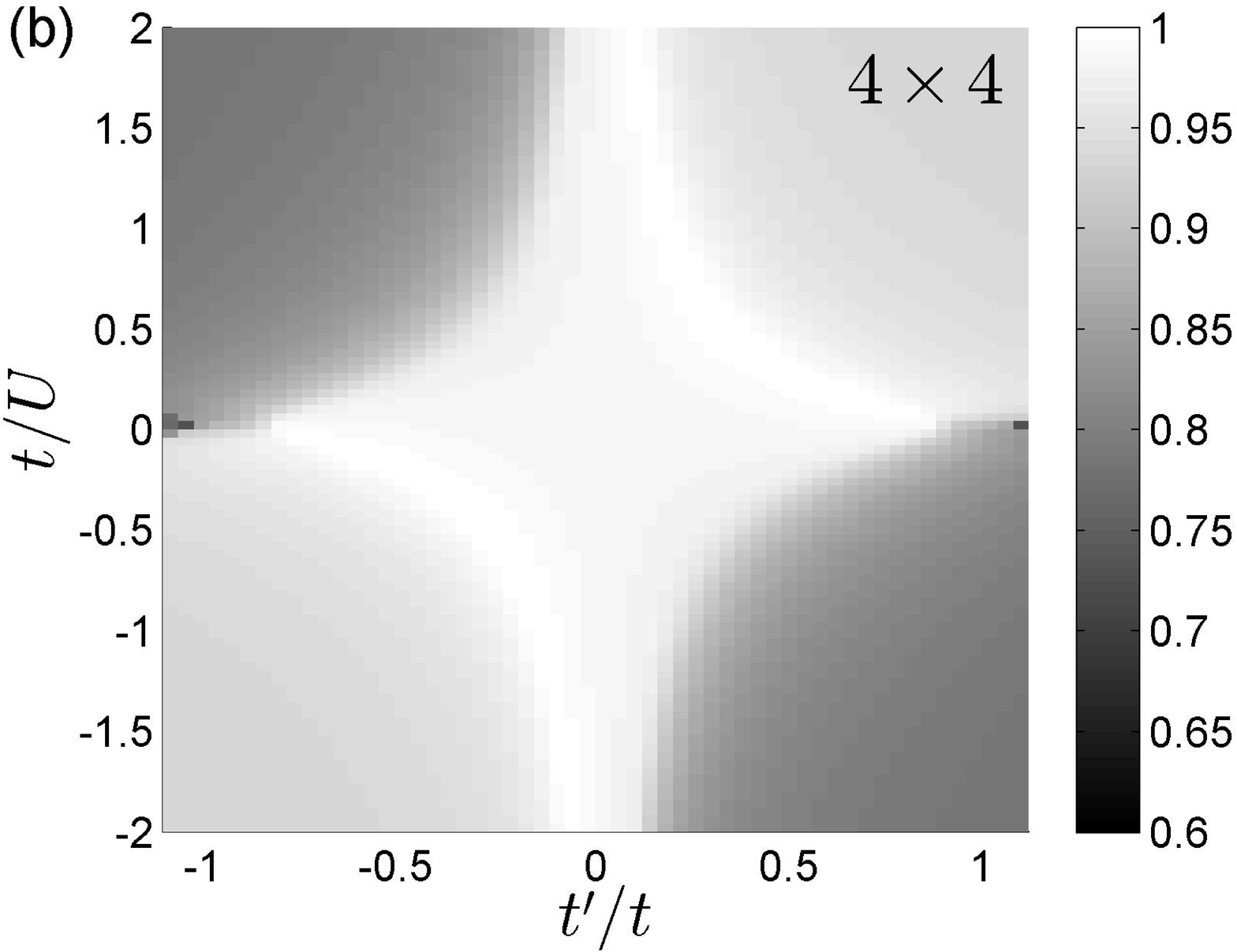}
\includegraphics[width=0.32\textwidth]{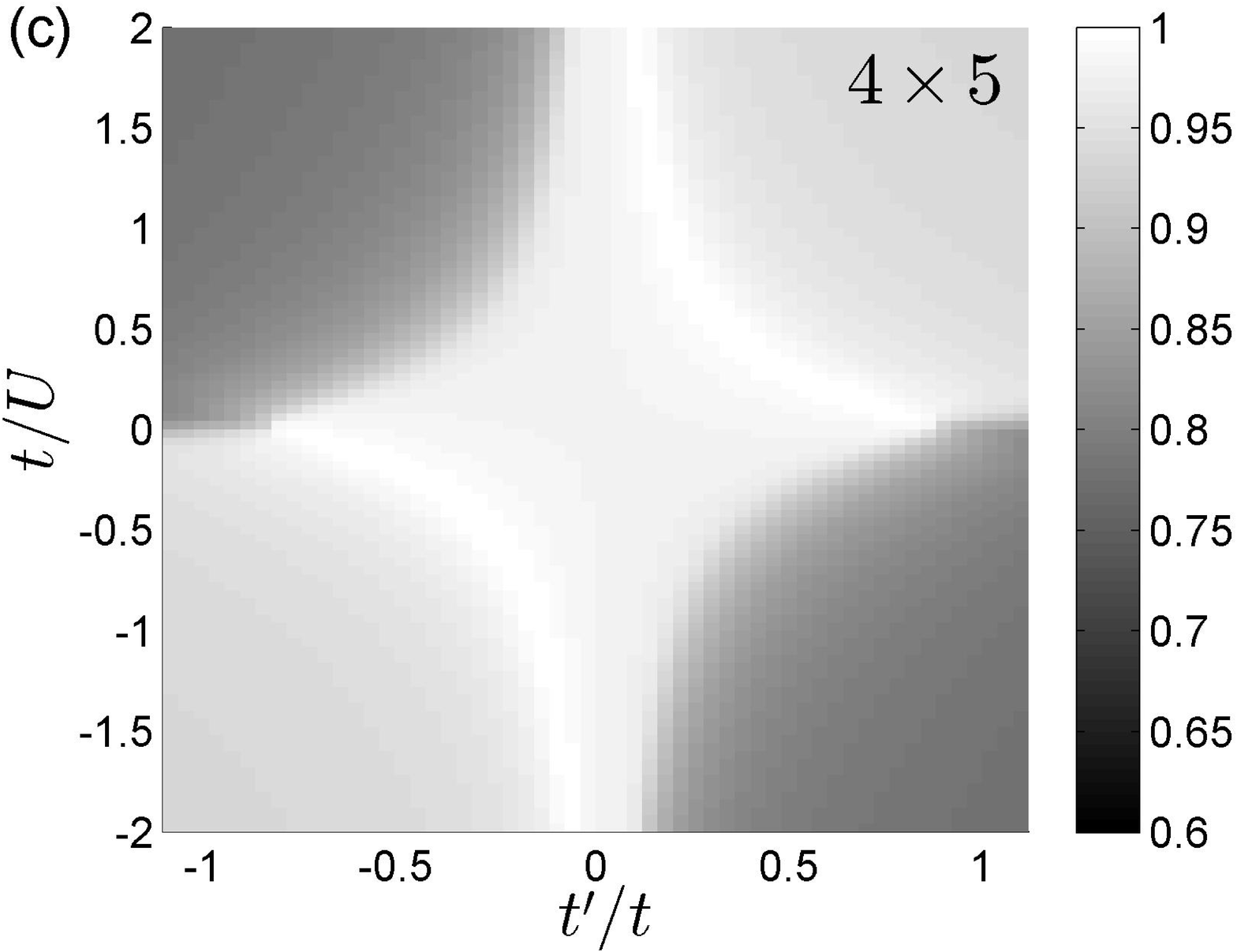}
\caption{Overlap per site $|\langle\psi_0|\psi\rangle|^{1/N}$ between the lowest energy eigenstate $\psi_0$ of the Hamiltonian in \eqref{Heff} (with $U>0$) and the FQH-like state $\psi$ in \eqref{CFT} as a function of $t/U$ and $t'/t$ for (a) a $4\times3$ lattice, (b) a $4\times4$ lattice, and (c) a $4\times5$ lattice. The Schrieffer-Wolff transformation leading from \eqref{FH} to \eqref{Heff} is valid when $|t|/U\ll1$ and $|t'|/U\ll1$, and we note that part of the area displaying an almost perfect overlap is within this region.}\label{fig:overlap}
\end{figure*}

The derivation of the effective Hamiltonian is done in Appendix \ref{sec:effmodel} and to third order in $t/U$ and $t'/U$ it gives
\begin{multline}\label{Heff}
H_{\rm eff}=\frac{2t^2}{U}\sum_{<n,m>} \left(2\vec{S}_n\cdot\vec{S}_m+\frac{1}{2}\right)\\
+\frac{2t'^2}{U}\sum_{<\!\!<n,m>\!\!>} \left(2\vec{S}_n\cdot\vec{S}_m+\frac{1}{2}\right)\\
-\frac{6t^2t'}{U^2}\sum_{<n,m,p>_{\circlearrowleft}}
4\vec{S}_n\cdot\left(\vec{S}_m\times\vec{S}_p\right)+{\rm constant}.
\end{multline}
The first sum is over all pairs of nearest neighbors, the second sum is over all pairs of next-nearest neighbors, and the third sum is over all triangles that are lattice translations of one of the four triangles marked in Fig.~\ref{fig:Hkin}. Each triangle is included only once and $n$, $m$, $p$ label the lattice sites at the vertices of the triangle in the counter clockwise direction as indicated with the arrow ${\circlearrowleft}$.

The action of the terms in $H_{\rm eff}$ is to permute spins since
\begin{equation}
\left(2\vec{S}_n\cdot\vec{S}_m+\frac{1}{2}\right)|\ldots \sigma_n\ldots \sigma_m\ldots\rangle=|\ldots \sigma_m\ldots \sigma_n\ldots\rangle
\end{equation}
and
\begin{multline}
 4\vec{S}_n\cdot(\vec{S}_m\times\vec{S}_p) |\ldots\sigma_n\ldots\sigma_m\ldots\sigma_p\ldots\rangle\\
=i|\ldots\sigma_p\ldots\sigma_n\ldots\sigma_m\ldots\rangle -i|\ldots\sigma_m\ldots\sigma_p\ldots\sigma_n\ldots\rangle.
\end{multline}
Therefore $H_{\rm eff}$ is also $SU(2)$ invariant, as it should be. The three-body term in \eqref{Heff} breaks time reversal symmetry, and we note that the chirality is build into the model by following the rule given in Fig.~\ref{fig:Hkin} for choosing the phases of the hopping amplitudes.

The Hamiltonian $H_{\rm eff}$ can be seen as a short-range version of the Hamiltonian presented in \cite{nielsen2}. The latter is an exact parent Hamiltonian for the state
\begin{multline}\label{CFT}
\psi(s_1,s_2,\ldots,s_N)=
\delta_\mathbf{s}\prod_{n=1}^N(-1)^{(n-1)(s_n+1)/2}\\ \times\prod_{n<m}\left(z_n-z_m\right)^{(s_ns_m+1)/2},
\end{multline}
where $s_n=+1$ ($-1$) when $\sigma_n={\uparrow}$ (${\downarrow}$), $z_n$ is the position of lattice site number $n$ written as a complex number, and $\delta_\mathbf{s}=1$ for $\sum_ns_n=0$ and $\delta_\mathbf{s}=0$ otherwise. As discussed in \cite{nielsen2,nielsen3,tu2}, \eqref{CFT} is a slightly modified version of the Kalmeyer-Laughlin state \cite{KL1,KL2}, which, up to some phase factors, is the $\nu=1/2$ Laughlin state with the possible particle positions limited to the sites of a square (or triangular) lattice. In fact, \eqref{CFT} reduces exactly to the Kalmeyer-Laughlin state in the thermodynamic limit \cite{nielsen2,tu2}. Several topological properties of \eqref{CFT} have been analyzed in \cite{nielsen3} and are in agreement with those of the $\nu=1/2$ Laughlin state in the continuum.

For small systems, the ground state $\psi_0$ of the Hamiltonian \eqref{Heff} can be obtained from exact diagonalization \cite{laeuchli}. In doing so, we use the conservation of the total spin in the $z$-direction, the symmetry under simultaneous rotation of all the spins by $180^\circ$ around the $x$-axis, and the symmetry under rotation of the lattice by $180^\circ$ to rewrite the Hamiltonian into block diagonal form, which reduces the size of the matrices that need to be diagonalized. In Fig.~\ref{fig:overlap}, we compare $\psi_0$ to the wave function \eqref{CFT} by computing the overlap per site $|\langle\psi_0|\psi\rangle|^{1/N}$ for different lattice sizes. We use here the overlap per site rather than the overlap because the overlap per site is more suitable for comparing results obtained for different lattice sizes. This is because the overlap generally decreases exponentially with system size in many-body systems due to the exponential increase in Hilbert space dimension, and the exponent $1/N$ appearing in the overlap per site counteracts this effect. The figure shows that the overlap per site is very close to unity for appropriately chosen parameters, and that the results are similar for all of the considered lattices. The region, where the overlap per site is high, also includes parameters with $|t|/U$ and $|t'|/U$ small. It is thus possible to create the state \eqref{CFT} with high fidelity by implementing the Fermi-Hubbard-like Hamiltonian in \eqref{FH} for appropriate parameters.

To further establish the connection to the bosonic Laughlin state at half filling, one could also look for gapless edge excitations in the spectrum of \eqref{Heff} in the thermodynamic limit. The system sizes that we can investigate with exact diagonalization for open boundary conditions are, however, too small to draw conclusions about the presence or absence of such states.

\subsection{Properties of $H_{{\rm kin},\sigma}$}

The Fermi-Hubbard-like Hamiltonian in \eqref{FH} consists of two kinetic energy terms describing free fermions hopping on a lattice and an interaction term, and one may therefore ask if the model can be seen as a flat band model with a partially filled and very flat energy band with nonzero Chern number plus interactions as described in the introduction. To investigate this question, we compute the band filling, the flatness parameter, and the Chern number for $H_{{\rm kin},\sigma}$ in the following. As these quantities are properties of the band structure, we shall consider periodic boundary conditions in this section and take the limit of an infinite lattice.

The band structure of $H_{{\rm kin},\sigma}$ consists of two bands because $H_{{\rm kin},\sigma}$ is periodic with period two lattice constants in the $x$-direction and one lattice constant in the $y$-direction as can be seen by inspection of Fig.~\ref{fig:Hkin}. To compute the band structure, we use a slightly modified notation, in which $a_{n,m,\sigma}$ is the annihilation operator of a fermion with spin $\sigma$ on lattice site $(n,m)$ with $n=1,2,\ldots,L_x$ and $m=1,2,\ldots,L_y$ and define the momentum space annihilation operators
\begin{align}
a_{p,q,\sigma}&=\sqrt{\frac{2}{N}}\sum_{n=1}^{L_x/2}\sum_{m=1}^{L_y}a_{2n,m,\sigma}
e^{-i\frac{4\pi}{L_x}pn}e^{-i\frac{2\pi}{L_y}qm}, \\
b_{p,q,\sigma}&=\sqrt{\frac{2}{N}}\sum_{n=1}^{L_x/2}\sum_{m=1}^{L_y}a_{2n-1,m,\sigma}
e^{-i\frac{4\pi}{L_x}pn}e^{-i\frac{2\pi}{L_y}qm}.
\end{align}
Reexpressing $H_{{\rm kin},\sigma}$ in terms of these, we get
\begin{equation}
H_{{\rm kin},\sigma}=\sum_{p=1}^{L_x/2}\sum_{q=1}^{L_y}
\left(\begin{array}{cc}a_{p,q,\sigma}^\dag & b_{p,q,\sigma}^\dag\end{array}\right)
\mathcal{H}_{pq}
\left(\begin{array}{c}a_{p,q,\sigma} \\ b_{p,q,\sigma}\end{array}\right),
\end{equation}
where
\begin{widetext}
\begin{equation}
\mathcal{H}_{pq}\!=\!\!
\left[\!\!\!\begin{array}{cc}2t\sin(\frac{2\pi q}{L_y}) & \hspace{-3.5mm}i t(e^{i\frac{4\pi}{L_x}p}-1)+i2t'(e^{i\frac{4\pi}{L_x}p}+1) \cos(\frac{2\pi q}{L_y})\\
-i t(e^{-i\frac{4\pi}{L_x}p}-1)
-i2t'(e^{-i\frac{4\pi}{L_x}p}+1)\cos(\frac{2\pi q}{L_y}) & \hspace{-3.5mm} -2t\sin(\frac{2\pi q}{L_y}) \end{array}\!\!\!\right]\!.
\end{equation}
Let $k_x=4\pi p/L_x$ and $k_y=2\pi q/L_y$. The eigenvalues of $\mathcal{H}_{pq}$ then take the form
\begin{equation}
\lambda_{\pm,k_x,k_y}=\pm\sqrt{4t^2\sin^2(k_y)
+2t^2(1-\cos(k_x))+8t'^2(1+\cos(k_x))\cos^2(k_y)}.
\end{equation}
\end{widetext}
The bands do hence not overlap, and the $N/2$ fermions with spin $\sigma$ precisely fill the lowest of the two bands.

\begin{figure}
\includegraphics[width=0.96\columnwidth]{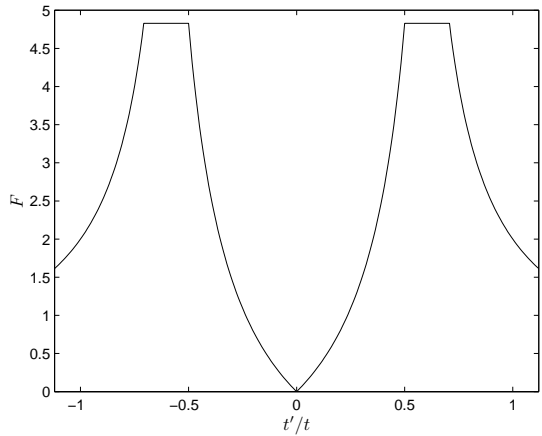}
\caption{Flatness $F$ [see \eqref{Fdef}] of the lowest energy band of $H_{{\rm kin},\sigma}$ [see \eqref{Hkin}] as a function of the ratio of the two hopping strengths $t'$ and $t$.} \label{fig:flat}
\end{figure}

The flatness parameter is defined as
\begin{equation}\label{Fdef}
F\equiv\frac{\min(E_{n+1})-\max(E_n)} {\max(E_n)-\min(E_n)},
\end{equation}
where $E_n$ is the set of energies of the $n$th energy band and band number $n$ is the highest energy band that is not completely empty at zero temperature. Since $\lambda_{+,k_x,k_y}=-\lambda_{-,k_x,k_y}\geq0$, we get in our case
\begin{equation}
F=\frac{2\min(\lambda_{+,k_x,k_y})}{\max(\lambda_{+,k_x,k_y})-\min(\lambda_{+,k_x,k_y})},
\end{equation}
where the maximum and minimum are with respect to $k_x$ and $k_y$. The extrema of $\lambda_{+,k_x,k_y}$ can be derived analytically, and from this we get
\begin{equation}\label{F}
F=\left\{\begin{array}{cl}
\frac{2\sqrt{2}t'}{t-\sqrt{2}t'} & {\rm for\ }\frac{t'}{t} \in\;\left[0,\frac{1}{2}\right[\\
\frac{2}{\sqrt{2}-1} & {\rm for\ }\frac{t'}{t} \in\;\left[\frac{1}{2},\frac{1}{\sqrt{2}}\right]\\
\frac{2t}{2t'-t} & {\rm for\ }\frac{t'}{t} \in\;\left]\frac{1}{\sqrt{2}},\infty\right[
\end{array}\right..
\end{equation}
The constant value of $F$ for $t'/t$ between $1/2$ and $1/\sqrt{2}$ appears because $\max(\lambda_{+,k_x,k_y})$ and $\min(\lambda_{+,k_x,k_y})$ depend on $t'/t$ through the same $(t'/t)$-dependent factor in this interval. Note that $F$ is symmetric around $t'/t=0$ since $\lambda_{+,k_x,k_y}$ does not depend on the signs of $t$ and $t'$. It follows that the flatness, which is plotted in Fig.~\ref{fig:flat}, never exceeds $2/(\sqrt{2}-1)\approx4.83$. Using (12) in \cite{shi}, we find numerically that the Chern number of the lowest energy band is plus or minus one for all nonzero $t$ and $t'$.

The above results show that the band filling in the free fermion model $H_{{\rm kin},\sigma}$ is not fractional and that the flatness is moderate compared to flat band models that may have flatness 20-50 \cite{sheng}, even though a flatness of 5 or 7 may also suffice \cite{neupert}. In addition, we use fermions and not bosons to implement a lattice version of a bosonic FQH state. The Fermi-Hubbard-like model is thus not of the flat band type, but nevertheless leads to a FQH state on a lattice.

\section{Implementation}\label{sec:implementation}

Before going into the details of the proposed implementation scheme, we give here a brief summary explaining the main ideas. Let us first briefly recall the setting used to simulate the standard Fermi-Hubbard model with real hopping amplitudes on a square lattice in ultracold fermionic atoms in optical lattices \cite{kohl,jordens,schneider,esslinger}. The first ingredient is to create an optical lattice from counter propagating laser beams. The interference between the laser fields gives rise to an intensity pattern that varies sinusoidally in space, and for the atoms this translates into a potential landscape of the same shape if the frequency is chosen appropriately. At low temperatures the atoms are trapped at the potential minima, which form a square lattice. If the difference between the potential minima and maxima is not too large, there is a non-negligible probability for an atom to tunnel through the potential barrier between two sites. This gives rise to the kinetic energy terms in the Hubbard model. If two atoms sit on the same site, their wave functions overlap, and they interact with each other. This gives rise to the on-site interaction terms.

To implement the Fermi-Hubbard-like model in \eqref{FH} we need some modifications of the above approach. First, we need to be able to print spatially varying phase factors on the tunneling amplitudes between nearest neighbor sites, and second, we need to also have tunneling between next-nearest neighbor sites. We propose to achieve this in the following way. We encode the spin degree of freedom in four internal hyperfine states of the fermions. We shall refer to these four states as the blue spin up state, the blue spin down state, the red spin up state, and the red spin down state, respectively. We would like the blue states and the red states to see two different potential landscapes, and we shall refer to these as the blue and the red potential, respectively. This can be achieved by choosing the hyperfine states such that the blue (red) states interact more strongly with right (left) circularly polarized light than with left (right) circularly polarized light, and then create different intensity patterns in space for the two polarizations.

\begin{figure}
\includegraphics[width=\columnwidth]{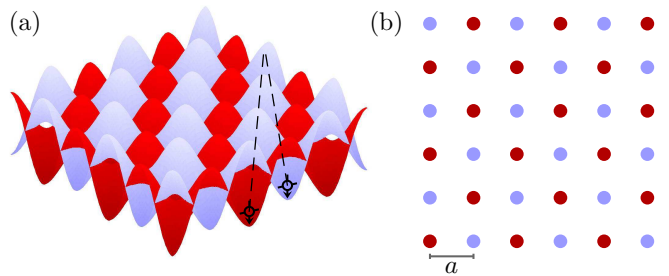}
\caption{(Color online) Checkerboard optical lattice potential. Fermions in the red spin up or in the red spin down state see the red potential in (a) and are hence trapped on the lattice sites that belong to the red sublattice shown in (b). Fermions in the blue spin up or in the blue spin down state see the blue potential in (a) and are hence trapped on the blue lattice sites in (b). A hop between blue and red lattice sites can be accomplished via a Raman transition that changes the internal state of the fermion as illustrated schematically in part (a) of the figure.} \label{fig:optlat}
\end{figure}

In particular, we would like the potential seen by the blue (red) states to have minima at the white (black) squares of a checkerboard as illustrated in Fig.~\ref{fig:optlat}. In this setting, tunneling events can happen between next-nearest neighbor sites on the lattice, since this corresponds to tunneling events between nearest neighbor minima in either the blue or the red potential. To move an atom between nearest-neighbor sites on the lattice, on the other hand, we need to also change the internal state of the atom. This relation between internal state and position allows us to use laser assisted tunneling \cite{jaksch} to implement the nearest neighbor hopping terms. The idea is to use two laser fields to drive Raman transitions between internal states with different color labels but the same spin labels. Since the potential energy would change drastically if the atom stayed at the same site during the transition, it can be forced to hop to a neighboring site during the transition by choosing the laser frequencies appropriately. This is also illustrated in Fig.~\ref{fig:optlat}. The advantage of implementing the nearest neighbor hopping terms in this way is that the relative phase of the two Raman lasers is printed on the hopping amplitudes. As we shall see below, it is possible to choose the spatial variation of the phases of the lasers in such a way that the desired phases on the hopping amplitudes are obtained. The lasers used for the Raman transitions also give a second contribution to the next-nearest neighbor hopping terms. Finally, the on-site interaction terms appear in the same way as for the standard Fermi-Hubbard model implementation.

\begin{figure}
\includegraphics[width=0.58\columnwidth]{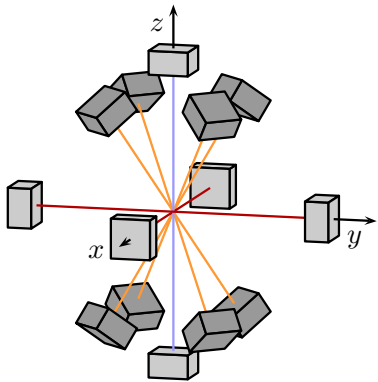}
\caption{(Color online) Schematic illustration of the laser beams needed for the implementation. The laser beams originating from the eight dark gray boxes produce the checkerboard optical lattice in the $xy$-plane, and the laser beams originating from the six light gray boxes implement the hopping terms and take care of trapping in the $z$-direction.} \label{fig:lasers}
\end{figure}

Let us finally give an overview of the needed laser configuration. As we shall show in the more detailed sections below, the desired checkerboard optical lattice can be created by eight laser beams coming from skew directions, and the laser assisted hopping terms and the trapping in the direction perpendicular to the plane of the lattice can be implemented with three standing wave laser fields along the axes of the setup. This is illustrated very schematically in Fig.~\ref{fig:lasers}.

\subsection{Checkerboard optical lattice potential}

We now turn to a detailed explanation of how one can create the optical lattice potential in Fig.~\ref{fig:optlat}(a).

\subsubsection{Atomic levels}

\begin{figure}
\includegraphics[width=0.725\columnwidth]{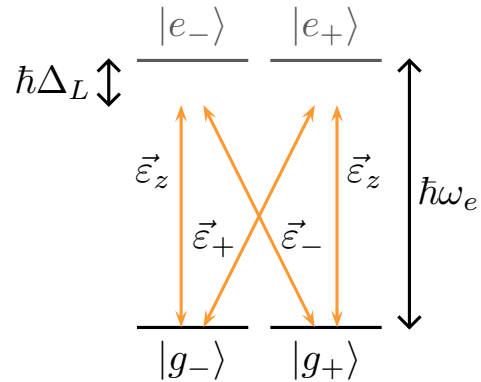}
\caption{(Color online) The atoms are assumed to have a $^2S_{1/2}$ ground state manifold and a $^2P_{1/2}$ excited state manifold, which gives the fine structure states shown in the figure. The drawing also displays the transitions driven by off-resonant left and right circularly polarized light and linearly $z$-polarized light, respectively.}\label{fig:levels}
\end{figure}

We shall here consider the case, where the ground state manifold of the atoms has term symbol $^2S_{1/2}$ and the trapping laser fields couple the ground states off-resonantly to a $^2P_{1/2}$ excited state manifold. This situation can be achieved with alkali atoms. Since the light fields do not interact with the spin of the nuclei of the atoms, and since we shall assume that the detuning is much larger than the hyperfine splitting, it is sufficient at this stage to consider the fine structure displayed in Fig.~\ref{fig:levels}. The ground state manifold then consists of the states
\begin{align}
|g_+\rangle&\equiv|0,1/2,1/2\rangle_J=|0,0,1/2\rangle_{LS},\\
|g_-\rangle&\equiv|0,1/2,-1/2\rangle_J=|0,0,-1/2\rangle_{LS}.\nonumber
\end{align}
The kets with subscript $J$ give the state in the form $|L,J,m_J\rangle$ and the kets with subscript $LS$ give the state in the form $|L,m_L,m_S\rangle$. Here, $S$ is the spin, $L$ is the orbital angular momentum, $J$ is the momentum obtained by coupling $S$ and $L$, and $m_S$, $m_L$, and $m_J$ are the $z$-components of the angular momenta \cite{sakurai}. The excited state manifold consists of the two states
\begin{align}
\left|e_+\right>&\equiv
\left|1,\frac{1}{2},\frac{1}{2}\right\rangle_J
=\sqrt{\frac{1}{3}}\left|1,0,\frac{1}{2}\right\rangle_{LS}\nonumber\\
&\hspace{35mm}-\sqrt{\frac{2}{3}}\left|1,1,-\frac{1}{2}\right\rangle_{LS},\\
\left|e_-\right>&\equiv
\left|1,\frac{1}{2},-\frac{1}{2}\right\rangle_J
=\sqrt{\frac{2}{3}}\left|1,-1,\frac{1}{2}\right\rangle_{LS}\nonumber\\
&\hspace{35mm}-\sqrt{\frac{1}{3}}\left|1,0,-\frac{1}{2}\right\rangle_{LS}.\nonumber
\end{align}
In the following, we denote the energy of $|e_\pm\rangle$ relative to the energy of $|g_\pm\rangle$ by $\hbar\omega_e$.

\subsubsection{Light fields}

The light fields needed to implement the optical lattice are summarized in table~\ref{tab:optlat}. The first row, e.g., represents the field
\begin{multline}\label{E1a}
\vec{E}_{1a}(\vec{r},t)
={\rm Re}\Big\{Ee^{i k_x(x+y)+i k_zz}e^{-i\omega_1 t}\\
\times\Big[\vec{\varepsilon}_+ +\alpha i\vec{\varepsilon}_-
+\frac{1}{\sqrt{2}\beta}(1+i)(1-\alpha)\vec{\varepsilon}_z\Big]\Big\},
\end{multline}
where
\begin{align}\label{kxkz}
k_x&=\frac{k}{\sqrt{2+\beta^2}}, &
k_z&=\frac{k\beta}{\sqrt{2+\beta^2}}, &
k&=\frac{2\pi}{\lambda},
\end{align}
$\lambda$ is the wavelength of the fields, $\vec{\varepsilon}_+$ is the polarization vector of right circularly polarized light, $\vec{\varepsilon}_-$ is the polarization vector of left circularly polarized light, and $\vec{\varepsilon}_z$ is the polarization vector of linearly $z$-polarized light, i.e.,
\begin{align}\label{polvec}
\vec{\varepsilon}_\pm&=\frac{1}{\sqrt{2}}\left(\begin{array}{c} \mp1 \\ -i\\ 0\end{array}\right),  &
\vec{\varepsilon}_z&=\left(\begin{array}{c} 0 \\ 0\\ 1\end{array}\right).
\end{align}

\begin{table*}
\begin{ruledtabular}
\begin{tabular}{@{}llll@{}}
Label & Frequency & Wave vector & Amplitude $\times$ Polarization \\
\hline
$1a$ & $\omega_1$ & $+k_x\hat{x}+k_x\hat{y}+k_z\hat{z}$ & $E\vec{\varepsilon}_++\alpha Ei\vec{\varepsilon}_-
+\frac{E}{\sqrt{2}\beta}(1+i)(1-\alpha)\vec{\varepsilon}_z$ \\
$1b$ & $\omega_1$ & $-k_x\hat{x}-k_x\hat{y}+k_z\hat{z}$ & $E\vec{\varepsilon}_+-\alpha Ei\vec{\varepsilon}_-
-\frac{E}{\sqrt{2}\beta}(1+i)(1+\alpha)\vec{\varepsilon}_z$ \\
$2a$ & $\omega_2$ & $+k_x\hat{x}+k_x\hat{y}-k_z\hat{z}$ & $E\vec{\varepsilon}_++\alpha Ei\vec{\varepsilon}_-
-\frac{E}{\sqrt{2}\beta}(1+i)(1-\alpha)\vec{\varepsilon}_z$ \\
$2b$ & $\omega_2$ & $-k_x\hat{x}-k_x\hat{y}-k_z\hat{z}$ & $E\vec{\varepsilon}_+-\alpha Ei\vec{\varepsilon}_-
+\frac{E}{\sqrt{2}\beta}(1+i)(1+\alpha)\vec{\varepsilon}_z$ \\
$3a$ & $\omega_3$ & $-k_x\hat{x}+k_x\hat{y}+k_z\hat{z}$ & $E\vec{\varepsilon}_++\alpha Ei\vec{\varepsilon}_-
-\frac{E}{\sqrt{2}\beta}(1-i)(1+\alpha)\vec{\varepsilon}_z$ \\
$3b$ & $\omega_3$ & $+k_x\hat{x}-k_x\hat{y}+k_z\hat{z}$ & $E\vec{\varepsilon}_+-\alpha Ei\vec{\varepsilon}_-
+\frac{E}{\sqrt{2}\beta}(1-i)(1-\alpha)\vec{\varepsilon}_z$ \\
$4a$ & $\omega_4$ & $-k_x\hat{x}+k_x\hat{y}-k_z\hat{z}$ & $E\vec{\varepsilon}_++\alpha Ei\vec{\varepsilon}_-
+\frac{E}{\sqrt{2}\beta}(1-i)(1+\alpha)\vec{\varepsilon}_z$ \\
$4b$ & $\omega_4$ & $+k_x\hat{x}-k_x\hat{y}-k_z\hat{z}$ & $E\vec{\varepsilon}_+-\alpha Ei\vec{\varepsilon}_-
-\frac{E}{\sqrt{2}\beta}(1-i)(1-\alpha)\vec{\varepsilon}_z$
\end{tabular}
\end{ruledtabular}
\caption{The eight light fields used to produce the checkerboard optical lattice. Here, $k_x$ and $k_z$ are defined in \eqref{kxkz}, $\hat{x}$, $\hat{y}$, and $\hat{z}$ are unit vectors in the $x$-, $y$-, and $z$-directions, respectively, $E$ is a complex number that adjusts the amplitudes of the fields, $\vec{\varepsilon}_\pm$ and $\vec{\varepsilon}_z$ are polarization vectors defined in \eqref{polvec}, and $\alpha$ and $\beta$ are real, adjustable parameters. The frequencies $\omega_1$, $\omega_2$, $\omega_3$, and $\omega_4$ are almost the same but differ sufficiently to ensure that there is no coherent interference. Note that the required differences are small enough that the lengths of the wave vectors are practically the same for all the beams.} \label{tab:optlat}
\end{table*}

Let us first consider the fields in the first two rows of the table. The wave vectors are chosen such that the fields produce a standing wave pattern along the $(x+y)$-direction and both fields have the same $z$-component of the wave vector. The $z$-component must be nonzero in order to be able to choose different phases and amplitudes of the left and right circularly polarized components of the fields. Specifically, the phases of the numbers multiplying $\vec{\varepsilon}_+$ and $\vec{\varepsilon}_-$ in the last column of the table are chosen such that the spatial variation in the $xy$-plane after adding the two fields is given by $\cos[k_x(x+y)]$ for the right circularly polarized component and by $\sin[k_x(x+y)]$ for the left circularly polarized component. This ensures that the intensity maxima of the right circularly polarized component are displaced relative to the intensity maxima of the left circularly polarized component as desired. The $\alpha$ is included to be able to adjust the relative strengths of the left and right circularly polarized components. Finally, the factors multiplying $\vec{\varepsilon}_z$ are fixed by the requirement that the wave vectors of the fields should be orthogonal to the polarization vectors.

As illustrated in Fig.~\ref{fig:levels}, the $\vec{\varepsilon}_\pm$ polarized component of the light interacts with atoms in the $|g_\mp\rangle$ state. If the field is red (blue) detuned, this interaction reduces (increases) the energy of $|g_\mp\rangle$ in regions of space, where the intensity of $\vec{\varepsilon}_\pm$ polarized light is high. This is what gives the desired trapping. The fact that there is also a $z$-polarized component of the field, however, leads to undesired Raman transitions between the states in the ground state manifold for the level structure we are considering as can be seen from Fig.~\ref{fig:levels}. This undesired effect can be canceled by adding the fields in the third and the fourth rows of the table. These fields are obtained from the first two by changing the sign of the $z$-component of both the wave vectors and the polarization vectors. The frequency is also changed slightly such that the second pair of fields do not interfere coherently with the first pair of fields. Note that this can be done without any significant change of the wavelength. The last four fields in the table similarly produce a standing wave pattern in the $(x-y)$-direction. The frequencies $\omega_3$ and $\omega_4$ differ slightly from $\omega_1$ and $\omega_2$ such that the fields do not interfere coherently.

Adding up all the fields in the table, we get the total electric field
\begin{equation}
\vec{E}(\vec{r},t)=\sum_{n=1}^4 \vec{E}_n(\vec{r},t)
\end{equation}
where
\begin{align}
\vec{E}_1(\vec{r},t)&={\rm Re}\left[2Ee^{i k_zz}e^{-i\omega_1 t}
\left(c_+\vec{\varepsilon}_+-\alpha s_+\vec{\varepsilon}_-
+d_+\vec{\varepsilon}_z\right)\right],\nonumber\\
\vec{E}_2(\vec{r},t)&={\rm Re}\left[2Ee^{-i k_zz}e^{-i\omega_2 t}
\left(c_+\vec{\varepsilon}_+-\alpha s_+\vec{\varepsilon}_- -d_+\vec{\varepsilon}_z\right)\right],\nonumber\\
\vec{E}_3(\vec{r},t)&={\rm Re}\left[2Ee^{i k_zz}e^{-i\omega_3 t} \left(c_-\vec{\varepsilon}_++\alpha s_-\vec{\varepsilon}_- +d_-\vec{\varepsilon}_z\right)\right],\nonumber\\
\vec{E}_4(\vec{r},t)&={\rm Re}\left[2Ee^{-i k_zz}e^{-i\omega_4 t} \left(c_-\vec{\varepsilon}_++\alpha s_-\vec{\varepsilon}_-
-d_-\vec{\varepsilon}_z\right)\right],\nonumber
\end{align}
and
\begin{align}
 c_\pm&=\cos\left[k_x(x\pm y)\right], \quad
s_\pm=\sin\left[k_x(x\pm y)\right], \\
d_\pm&=\frac{1}{\sqrt{2}\beta}(1\pm i)(i s_\pm-\alpha c_\pm).\nonumber
\end{align}

Let us finally note that a small geometric consideration shows that the wavelength of the light fields is related to $\beta^2$ through the relation
\begin{equation}
\lambda=\frac{4a}{\sqrt{2+\beta^2}},
\end{equation}
where $a$ is the lattice constant as illustrated in Fig.~\ref{fig:optlat}(b). Therefore the wavelength is approximately the same as the wavelength of the fields needed to induce the hopping terms if $\beta^2\approx2$. This is convenient because it is then possible to use the same set of excited states for the trapping and for the Raman transitions.

\subsubsection{Light-atom interaction}

Within the dipole approximation and ignoring all states that are not within the subspace spanned by $|g_\pm\rangle$ and $|e_\pm\rangle$, we can write the Hamiltonian of an atom interacting with the field $\vec{E}(\vec{r},t)$ as
\begin{multline}
\tilde{H}=\hbar\omega_e(|e_+\rangle\langle e_+|+|e_-\rangle\langle e_-|)\\
-(P_0+Q_0)\vec{d}\cdot\vec{E}(\vec{r},t)(P_0+Q_0),
\end{multline}
where $\vec{d}=-e\vec{\eta}$ is the dipole operator of the atom, $-e$ is the charge of an electron, $\vec{\eta}$ is the position of the electron interacting with the field with respect to the nucleus of the atom, $\vec{r}$ is the position of the nucleus, and
\begin{align}
P_0&=|g_+\rangle\langle g_+|+|g_-\rangle\langle g_-|,\\
Q_0&=|e_+\rangle\langle e_+|+|e_-\rangle\langle e_-|.\nonumber
\end{align}

We now move into a rotating frame defined by the Hamiltonian $H_{\rm RF}=\hbar\omega_1Q_0$. In this frame, the Hamiltonian of the system is
\begin{equation}
H=e^{i H_{\rm RF}t/\hbar}\tilde{H}e^{-i H_{\rm RF}t/\hbar}-H_{\rm RF},
\end{equation}
where
\begin{equation}
e^{i H_{\rm RF}t/\hbar}=e^{i\omega_1t}Q_0+P_0.
\end{equation}
Writing $H=H_0+V$, we get
\begin{equation}
H_0=-\hbar\Delta_L(|e_+\rangle\langle e_+|+|e_-\rangle\langle e_-|),
\end{equation}
with $\Delta_L=\omega_1-\omega_e$, and
\begin{multline}\label{Voptlat}
V=2eEQ_0\Big[e^{i k_zz}
\left(\vec{\eta}\cdot\vec{\varepsilon}_+c_+
-\alpha\vec{\eta}\cdot\vec{\varepsilon}_-s_+
+\vec{\eta}\cdot\vec{\varepsilon}_zd_+\right)\\
+e^{-i k_zz}e^{-i(\omega_2-\omega_1)t} \left(\vec{\eta}\cdot\vec{\varepsilon}_+c_+
-\alpha\vec{\eta}\cdot\vec{\varepsilon}_-s_+
-\vec{\eta}\cdot\vec{\varepsilon}_zd_+\right)\\
+e^{i k_zz}e^{-i(\omega_3-\omega_1)t} \left(\vec{\eta}\cdot\vec{\varepsilon}_+c_-
+\alpha\vec{\eta}\cdot\vec{\varepsilon}_-s_-
+\vec{\eta}\cdot\vec{\varepsilon}_zd_-\right)\\
+e^{-i k_zz}e^{-i(\omega_4-\omega_1)t} \left(\vec{\eta}\cdot\vec{\varepsilon}_+c_-
+\alpha\vec{\eta}\cdot\vec{\varepsilon}_-s_-
-\vec{\eta}\cdot\vec{\varepsilon}_zd_-\right)\Big]P_0\\
+h.c.,
\end{multline}
where $h.c.$ is the hermitian conjugate and we have used the rotating wave approximation to drop terms oscillating as $e^{\pm i (\omega_n+\omega_1) t}$ with $n=1,2,3,4$ and used $P_0\vec{\eta}P_0=Q_0\vec{\eta}Q_0=0$, which follows from the inversion symmetry of atoms. Note that $\Delta_L$ is negative for red detuning, which is typically what we shall consider.

The matrix elements of $\vec{\eta}\cdot\vec{\varepsilon}_\pm$ and $\vec{\eta}\cdot\vec{\varepsilon}_z$ can be computed by noting that the angular part of the atomic wavefunctions are spherical harmonics. Let $\psi_0(\eta)$ ($\psi_1(\eta)$) be the radial wavefunction of the states in the ground (excited) state manifold. The nonzero matrix elements are then
\begin{align}
\langle e_\pm|(\vec{\eta}\cdot\vec{\varepsilon}_z)|g_\pm\rangle&=\pm R/3, \\
\langle e_\pm|(\vec{\eta}\cdot\vec{\varepsilon}_\pm) |g_\mp\rangle&=\mp\sqrt{2}R/3,\nonumber
\end{align}
where
\begin{equation}\label{R}
R=\int_0^\infty \psi_1^*(\eta)\psi_0(\eta) \eta^3d\eta,
\end{equation}
and therefore
\begin{align}
Q_0(\vec{\eta}\cdot\vec{\varepsilon}_z)P_0&=
\frac{R}{3}|e_+\rangle\langle g_+|
-\frac{R}{3}|e_-\rangle\langle g_-|,\\
Q_0(\vec{\eta}\cdot\vec{\varepsilon}_\pm)P_0&=
\mp\frac{\sqrt{2}R}{3}|e_\pm\rangle\langle g_\mp|,\nonumber
\end{align}
which we insert into \eqref{Voptlat}.

\subsubsection{Effective Hamiltonian}

When $|\Delta_L|$ is large compared to the Rabi frequency, we can use the Schrieffer-Wolff transformation (see Appendix \ref{sec:SW}) to eliminate the excited states of the system. In this case, both the zeroth, first, and third order terms of the effective Hamiltonian \eqref{HeffSW} are zero, and the second order term simplifies such that
\begin{equation}\label{hs}
H_{\rm eff}=\frac{(Q_0VP_0)^\dag Q_0VP_0}{\hbar\Delta_L}.
\end{equation}
Since the four terms in the potential \eqref{Voptlat} have different frequencies, they can be treated independently, which amounts to dropping fast oscillating terms in \eqref{hs}. The contribution to $H_{\rm eff}$ coming from the term in the potential that is due to the fields in the first two rows of table~\ref{tab:optlat} is
\begin{multline}
 H_{{\rm eff},1}=\frac{4e^2|E|^2|R|^2}{9\hbar\Delta_L}
\Big[2|g_-\rangle\langle g_-|c_+^2+2\alpha^2|g_+\rangle\langle g_+|s_+^2\\
+|d_+|^2P_0
+\sqrt{2}|g_+\rangle\langle g_-|(-c_+d_+^*+\alpha s_+d_+)\\
+\sqrt{2}|g_-\rangle\langle g_+|(-c_+d_++\alpha s_+d_+^*)\Big].
\end{multline}
The fields in the third and the fourth row give the same contribution except that $d_+$ is changed to $-d_+$. Therefore the undesired terms giving rise to transitions between $|g_+\rangle$ and $|g_-\rangle$ are precisely canceled as claimed above.

Adding also the contributions from the last four fields, we get
\begin{multline}
H_{\rm eff}=\frac{16e^2|E|^2|R|^2}{9\hbar\Delta_L}
\Big[(c_+^2+c_-^2)|g_-\rangle\langle g_-|\\
+\alpha^2(s_+^2+s_-^2)|g_+\rangle\langle g_+|
+\frac{1}{2}\left(|d_+|^2+|d_-|^2\right)P_0\Big].
\end{multline}
Since
\begin{equation}
\frac{1}{2}(|d_+|^2+|d_-|^2)=
\frac{1}{2\beta ^2}[s_+^2+s_-^2+\alpha ^2(c_+^2+c_-^2)],
\end{equation}
we conclude that the potential energy landscape seen by a fermion in the state $|g_-\rangle$ is
\begin{equation}
V_-=-V_0[(2\beta ^2+\alpha ^2)(c_+^2+c_-^2)+(s_+^2+s_-^2)]
\end{equation}
and the potential energy landscape seen by a fermion in the state $|g_+\rangle$ is
\begin{equation}
V_+=-V_0[(1+2\alpha ^2\beta ^2)(s_+^2+s_-^2)+\alpha ^2(c_+^2+c_-^2)],
\end{equation}
where
\begin{equation}
V_0\equiv-\frac{8e^2|E|^2|R|^2}{9\hbar\Delta_L\beta^2}.
\end{equation}
Note that $V_0$ is positive for red detuning and negative for blue detuning.

\subsubsection{Lattice potentials for the hyperfine states}

As we shall later on consider laser fields that drive Raman transitions between different hyperfine levels, we shall now discuss the hyperfine structure. For simplicity we assume that the spin of the nucleus is $I=1$, which is, e.g., the case for $^6Li$ (see \cite{lithium}). Coupling the nuclear spin $I$ and the electron angular momentum $J$ to the total angular momentum $F$, one finds that the ground state manifold consists of the hyperfine levels \cite{sakurai}
\begin{align}\label{hfstates}
\left|0,\frac{3}{2},+\frac{3}{2}\right\rangle_F
&=\left|0,1,\frac{1}{2}\right\rangle_{IJ},\\
\left|0,\frac{3}{2},+\frac{1}{2}\right\rangle_F
&=\sqrt{\frac{2}{3}}\left|0,0,\frac{1}{2}\right\rangle_{IJ}
+\sqrt{\frac{1}{3}}\left|0,1,-\frac{1}{2}\right\rangle_{IJ}, \nonumber\\
\left|0,\frac{3}{2},-\frac{1}{2}\right\rangle_F
&=\sqrt{\frac{1}{3}}\left|0,-1,\frac{1}{2}\right\rangle_{IJ}
+\sqrt{\frac{2}{3}}\left|0,0,-\frac{1}{2}\right\rangle_{IJ}, \nonumber\\
\left|0,\frac{3}{2},-\frac{3}{2}\right\rangle_F
&=\left|0,-1,-\frac{1}{2}\right\rangle_{IJ}, \nonumber\\
\left|0,\frac{1}{2},+\frac{1}{2}\right\rangle_F
&=\sqrt{\frac{1}{3}}\left|0,0,\frac{1}{2}\right\rangle_{IJ}
-\sqrt{\frac{2}{3}}\left|0,1,-\frac{1}{2}\right\rangle_{IJ}, \nonumber\\
\left|0,\frac{1}{2},-\frac{1}{2}\right\rangle_F
&=\sqrt{\frac{2}{3}}\left|0,-1,\frac{1}{2}\right\rangle_{IJ}
-\sqrt{\frac{1}{3}}\left|0,0,-\frac{1}{2}\right\rangle_{IJ}\nonumber,
\end{align}
where kets with subscript $F$ give the state in the form $|L,F,m_F\rangle$, kets with subscript $IJ$ give the state in the form $|L,m_I,m_J\rangle$, and $m_F$, $m_I$, and $m_J$ are the $z$-components of the angular momenta.

\begin{figure}
\includegraphics[width=0.9\columnwidth]{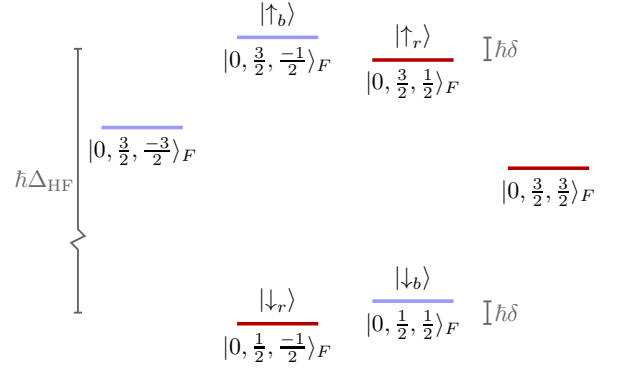}
\caption{(Color online) Hyperfine structure of the ground state manifold and the encoding of the red and blue spin up and down states. The states are labeled $|L,F,m_F\rangle$ as on the left hand side of \eqref{hfstates}. The vertical position of each state is the energy of the state in the checkerboard optical lattice when $\beta^2=2$, $\alpha^2=1.2$, and $V_0>0$. The states that see a potential with minima at the blue sites for these parameters are shown in blue and are assumed to be trapped on a blue site. The states that see a potential with minima at the red sites for these parameters are shown in red and are assumed to be trapped on a red site. Possible differences in zero point energy on the different lattice sites and the trapping in the $z$-direction are not taken into account in this drawing. Such effects will, however, not spoil the symmetry ensuring that the two $\hbar\delta$'s are the same. The hyperfine splitting $\hbar\Delta_{\rm HF}$ is not to scale.}\label{fig:hflevels}
\end{figure}

The potential energy landscapes
\begin{align}
V_{\frac{3}{2},\frac{3}{2}}&=V_+=-2V_0(1+2\alpha^2\beta^2)\nonumber\\
&\hspace{16mm} +V_0(1+2\alpha^2\beta^2-\alpha^2)(c_+^2+c_-^2),\label{Vp3}\\
V_{\frac{3}{2},\frac{1}{2}}&=V_{\frac{1}{2},-\frac{1}{2}}
=\frac{2}{3}V_++\frac{1}{3}V_-
=-2V_0\left(1+\frac{4}{3}\alpha^2\beta^2\right) \nonumber\\ &\hspace{1mm}+V_0\left(\frac{4}{3}\alpha^2\beta^2 -\alpha^2+1-\frac{2}{3}\beta^2\right)(c_+^2+c_-^2),\label{Vred}\\
V_{\frac{3}{2},-\frac{1}{2}}&=V_{\frac{1}{2},\frac{1}{2}}
=\frac{1}{3}V_++\frac{2}{3}V_-=-2V_0\left(\alpha^2+\frac{4}{3}\beta^2\right) \nonumber\\ &\hspace{1mm}+V_0\left(\alpha^2-\frac{2}{3}\alpha^2\beta^2+\frac{4}{3}\beta^2-1\right) (s_+^2+s_-^2),\label{Vblue}\\
V_{\frac{3}{2},-\frac{3}{2}}&=V_-=-2V_0(\alpha^2+2\beta^2)\nonumber\\ &\hspace{20mm}+V_0(\alpha^2+2\beta^2-1)(s_+^2+s_-^2).\label{Vm3}
\end{align}
seen by the hyperfine states are obtained as the diagonal matrix elements of $H_{\rm eff}$, i.e.\ $V_{F,m_F}={}_F\langle 0,F,m_F|H_{\rm eff}|0,F,m_F\rangle_F$. We note that the hyperfine splitting is not taken into account in $H_{\rm eff}$, but this splitting ensures that the fields do not induce transitions between hyperfine states with different values of $F$.

\begin{figure}
\includegraphics[width=\columnwidth]{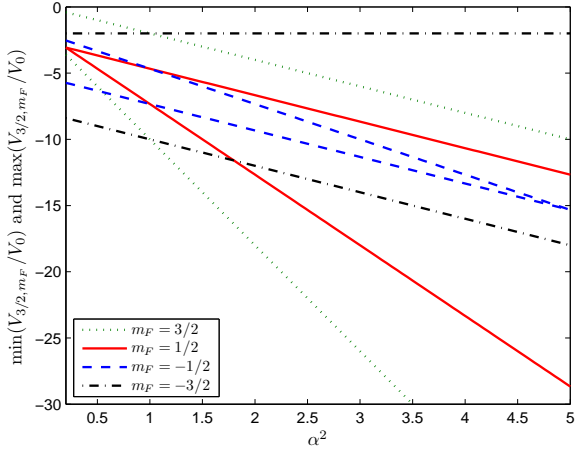}
\caption{(Color online) Minimum and maximum values of $V_{3/2,m_F}/V_0$ for $m_F=-3/2,-1/2,1/2,3/2$ and $\beta^2=2$ as a function of $\alpha^2$ in the region where \eqref{alphalim} is fulfilled.}\label{fig:potential}
\end{figure}

Assume we choose the parameters such that the coefficient of $(c_+^2+c_-^2)$ in \eqref{Vred} and of $(s_+^2+s_-^2)$ in \eqref{Vblue} are both positive. For red detuning ($V_0>0$) this is the case provided
\begin{equation}\label{alphalim}
\frac{\frac{2}{3}\beta^2-1}{\frac{4}{3}\beta^2-1}<\alpha^2
<\frac{\frac{4}{3}\beta^2-1}{\frac{2}{3}\beta^2-1}.
\end{equation}
We can then obtain the optical lattice potential in Fig.~\ref{fig:optlat} by implementing the red and blue spin up and down states such that
\begin{align}
&|{\uparrow}_r\rangle=|0,3/2,1/2\rangle_F, &
&|{\downarrow}_r\rangle=|0,1/2,-1/2\rangle_F, \\
&|{\uparrow}_b\rangle=|0,3/2,-1/2\rangle_F, &
&|{\downarrow}_b\rangle=|0,1/2,1/2\rangle_F,\nonumber
\end{align}
where the subscript $r$ ($b$) refers to red (blue). Figure~\ref{fig:hflevels} illustrates this encoding and the energy shifts due to the optical lattice potential. Note that $|{\uparrow}_r\rangle$ and $|{\downarrow}_r\rangle$ see exactly the same potential, and  $|{\uparrow}_b\rangle$ and $|{\downarrow}_b\rangle$ also see exactly the same potential.

In Fig.~\ref{fig:potential}, we plot the maximum and minimum values of $V_{F,m_F}$. The figure shows the freedom we have to adjust the relative heights of the red and blue potentials and the minimum of the red potential relative to the minimum of the blue potential by varying $\alpha^2$ when $\beta^2=2$.

\subsection{Implementation of the hopping terms}

\subsubsection{Light fields}

\begin{figure}
\includegraphics[width=0.83\columnwidth]{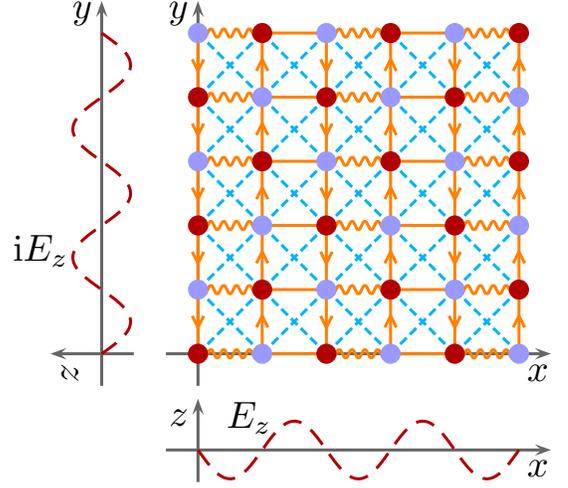}
\caption{(Color online) Implementation of the Fermi-Hubbard-like model \eqref{FH} with laser assisted tunneling on the checkerboard optical lattice. The kinetic energy part is implemented with the two $z$-polarized standing wave laser fields along, respectively, the $x$- and $y$-axis with frequency $\omega_r$ and a standing wave laser field (not shown) along the $z$-direction with frequency $\omega_b$ containing left and right circularly polarized components. Note that it is possible to use the same set of lasers to implement both $H_{{\rm kin},{\uparrow}}$ and $H_{{\rm kin},{\downarrow}}$, and that $H_{\rm int}$ comes from the interaction between two fermions with opposite spin sitting on the same site.}\label{fig:laasho}
\end{figure}

\begin{figure}
\includegraphics[width=0.8\columnwidth]{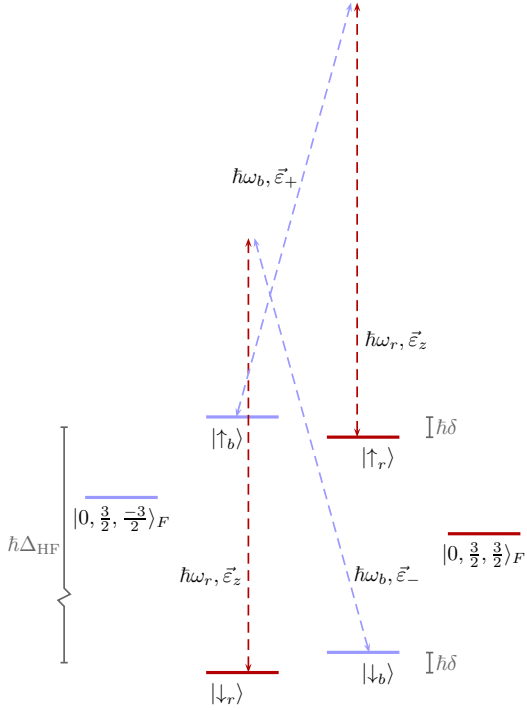}
\caption{(Color online) Raman transitions used for laser assisted tunneling between nearest-neighbor sites (the frequencies of the lasers are not to scale).}\label{fig:hf}
\end{figure}

To implement the hopping between red and blue lattice sites, we propose to use the standing wave fields
\begin{subequations}\label{fields}
\begin{align}
\vec{E}_{rx}&={\rm Re}(i E_z\vec{\varepsilon}_z e^{i kx}e^{-i\omega_r t}
-i E_z\vec{\varepsilon}_ze^{-i kx}e^{-i\omega_r t})\nonumber\\
&=-E_z\vec{\varepsilon}_z\sin(kx)e^{-i\omega_r t}+c.c.,\\
\vec{E}_{ry}&={\rm Re}(-E_z\vec{\varepsilon}_ze^{i ky}e^{-i\omega_r t} +E_z\vec{\varepsilon}_ze^{-iky}e^{-i\omega_r t})\nonumber\\
&=-i E_z\vec{\varepsilon}_z\sin(ky)e^{-i\omega_r t}+c.c.,\\
\vec{E}_{bz}&={\rm Re}[(E_+\vec{\varepsilon}_++E_-\vec{\varepsilon}_-)e^{i k_zz}e^{-i\omega_b t} \nonumber\\
&\hspace{18mm}+(E_+\vec{\varepsilon}_++E_-\vec{\varepsilon}_-)e^{-i k_zz}e^{-i\omega_b t}]\nonumber\\
&=(E_+\vec{\varepsilon}_++E_-\vec{\varepsilon}_-)\cos(k_zz)e^{-i\omega_b t}+c.c.,
\end{align}
\end{subequations}
created by three pairs of counter propagating laser beams along the $x$-, $y$-, and $z$-direction, respectively ($c.c.$ is the complex conjugate). We choose $k=\pi/a$ as illustrated in Fig.~\ref{fig:laasho}, where $a$ is the lattice constant. This also fixes $\omega_r$. The origin is assumed to be at the lower left corner of the lattice, and we choose $\omega_b$ such that $\omega_r-\omega_b=\delta$, where $\delta$ is the energy of an atom on a blue site minus the energy of an atom on a red site. Note that this energy difference is the same for spin up and spin down, and the fields \eqref{fields} therefore drive Raman transitions between both the up states and the down states as shown in Fig.~\ref{fig:hf}. If this symmetry is not present in a given setup, the field $\vec{E}_{bz}$ should be replaced by two fields with different frequencies and appropriate polarizations. Note also that the optical lattice automatically shifts away the energies of the states $|0,3/2,3/2\rangle_F$ and $|0,3/2,-3/2\rangle_F$ [see \eqref{Vp3} and \eqref{Vm3} and Fig.~\ref{fig:hf}] such that these states can be ignored in the following.

\subsubsection{Wannier functions}

Let us consider a lattice of infinite extent and denote the spatial part of the wave function of a fermion at the site with position coordinates $(x_n,y_n)$ by $f_b(x-x_n,y-y_n,z)$ if the site is blue and $f_r(x-x_n,y-y_n,z)$ if the site is red. Since we are considering only the lowest energy state in each of the potential wells of the optical lattice, we shall anticipate that $f_{b/r}(x,y,z)$ is symmetric under reflection in the $x$-axis, under reflection in the $y$-axis, and under reflection in the $z$-axis. Due to the symmetry of the lattice, we also assume that $f_{b/r}(x,y,z)$ is invariant under a rotation of $90^\circ$. For a deep lattice, e.g., $f_{b/r}(x,y,z)$ is a Gaussian in all three coordinates, but we shall not assume this in the following.

A fermion in the internal state $|\sigma_b\rangle$ (encoded in the blue spin up and down states) sitting on a blue site with coordinates $(x_n,y_n)$ is then described by the state
\begin{equation}
|\psi^{\sigma_b}_n\rangle=\int\!\!\!\int\!\!\!\int\! f_b(x-x_n,y-y_n,z) |x\rangle|y\rangle|z\rangle|\sigma_b\rangle d x d y d z.
\end{equation}
Likewise, a fermion in the internal state $|\sigma_r\rangle$ (encoded in the red spin up and down states) sitting on a red site with coordinates $(x_n,y_n)$ is described by the state
\begin{equation}
|\psi^{\sigma_r}_n\rangle=\int\!\!\!\int\!\!\!\int\! f_r(x-x_n,y-y_n,z) |x\rangle|y\rangle|z\rangle|\sigma_r\rangle d x d y d z.
\end{equation}
We shall also need expressions for the excited states used for the Raman transitions. We denote the internal part $|e_q\rangle$ and define $g_p(x,y,z)$, $p=1,2,\ldots$, to be a complete set of spatial wavefunctions. The states are then
\begin{equation}
|\psi_p^{e_q}\rangle=\int\!\!\!\int\!\!\!\int\! g_p(x,y,z) |x\rangle|y\rangle|z\rangle|e_q\rangle d x d y d z.
\end{equation}
Finally, for later convenience, we define the projectors
\begin{align}
&P_0=P_{r{\downarrow}}+P_{r{\uparrow}}+P_{b{\downarrow}}+P_{b{\uparrow}},\\
&Q_0=\sum_q\sum_p|\psi_p^{e_q}\rangle\langle\psi_p^{e_q}|,\nonumber\\
&P_{r{\downarrow}}=\sum_{n\in\mathcal{R}} |\psi_n^{{\downarrow}_r}\rangle\langle \psi_n^{{\downarrow}_r}|, \quad
P_{b{\downarrow}}=\sum_{n\in\mathcal{B}}|\psi_n^{{\downarrow}_b}\rangle\langle \psi_n^{{\downarrow}_b}|,\nonumber\\
&P_{r{\uparrow}}=\sum_{n\in\mathcal{R}}|\psi_n^{{\uparrow}_r}\rangle\langle \psi_n^{{\uparrow}_r}|, \quad
P_{b{\uparrow}}=\sum_{n\in\mathcal{B}}|\psi_n^{{\uparrow}_b}\rangle\langle \psi_n^{{\uparrow}_b}|,\nonumber
\end{align}
where $\mathcal{R}$ ($\mathcal{B}$) is the set of all the red (blue) lattice sites.

\subsubsection{Light-atom interaction}

In the dipole approximation and ignoring irrelevant levels, the Hamiltonian describing an atom in the optical lattice and its interaction with the standing wave fields is
\begin{multline}
\tilde{H}=\hbar\sum_{p,q}\omega_p^{e_q}|\psi_p^{e_q}\rangle\langle\psi_p^{e_q}| -\hbar\frac{\delta}{2}P_{r{\downarrow}}
+\hbar\left(\Delta_{\rm HF}-\frac{\delta}{2}\right)P_{r{\uparrow}}\\
+\hbar\frac{\delta}{2}P_{b{\downarrow}}
+\hbar\left(\Delta_{\rm HF}+\frac{\delta}{2}\right)P_{b{\uparrow}}
+e(P_0+Q_0)\vec{\eta}\cdot\vec{E}_T(P_0+Q_0),
\end{multline}
where $\vec{E}_T=\vec{E}_{rx}+\vec{E}_{ry}+\vec{E}_{bz}$ and $\hbar\Delta_{\rm HF}$ is the hyperfine energy splitting between the states in the ground state manifold with $F=3/2$ and $F=1/2$ as in Fig.~\ref{fig:hf}. We move into a rotating frame defined by the Hamiltonian
\begin{multline}
H_{\rm RF}=\hbar(\omega_r-\delta/2)Q_0
-\hbar\frac{\delta}{2}P_{r{\downarrow}}
+\hbar\left(\Delta_{\rm HF}-\frac{\delta}{2}\right)P_{r{\uparrow}}\\
+\hbar\frac{\delta}{2}P_{b{\downarrow}}
+\hbar\left(\Delta_{\rm HF}+\frac{\delta}{2}\right)P_{b{\uparrow}}.
\end{multline}
In this frame, the Hamiltonian of the system takes the form $H=H_0+V$, where
\begin{align}
H_0&=-\hbar\sum_{p,q}\Delta_{pq}|\psi_p^{e_q}\rangle\langle\psi_p^{e_q}|,\\
\Delta_{pq}&=\omega_r-\delta/2-\omega_p^{e_q},\nonumber
\end{align}
and
\begin{multline}
V=eQ_0(\vec{\eta}\cdot\vec{E}_T)P_{r{\downarrow}}e^{i\omega_rt}
+eQ_0(\vec{\eta}\cdot\vec{E}_T)P_{r{\uparrow}}e^{i(\omega_r-\Delta_{\rm HF})t}\\
+eQ_0(\vec{\eta}\cdot\vec{E}_T)P_{b{\downarrow}}e^{i\omega_bt}
+eQ_0(\vec{\eta}\cdot\vec{E}_T)P_{b{\uparrow}}e^{i(\omega_b-\Delta_{\rm HF})t}\\
+h.c..
\end{multline}
Inserting the fields \eqref{fields} and using the rotating wave approximation, we get
\begin{multline}
 V=\sum_{\sigma\in\{{\downarrow},{\uparrow}\}} \Big\{-eQ_0E_z\vec{\eta}\cdot\vec{\varepsilon}_z[\sin(kx)+i\sin(ky)]P_{r{\sigma}}\\
+eQ_0(E_+\vec{\eta}\cdot\vec{\varepsilon}_++E_-\vec{\eta}\cdot\vec{\varepsilon}_-) \cos(k_zz)P_{b{\sigma}}\\
-eQ_0E_z\vec{\eta}\cdot\vec{\varepsilon}_z[\sin(kx)+i\sin(ky)]P_{b{\sigma}}e^{-i\delta t}\\
+eQ_0(E_+\vec{\eta}\cdot\vec{\varepsilon}_++E_-\vec{\eta}\cdot\vec{\varepsilon}_-) \cos(k_zz)P_{r{\sigma}}e^{i\delta t}\Big\}
e^{-i\delta_{\sigma{\uparrow}}\Delta_{\rm HF}t}\\
+h.c.,
\end{multline}
where $\delta_{\sigma{\uparrow}}$ is a Kronecker delta function.

\subsubsection{Effective Hamiltonian and hopping amplitudes}

Assuming that $|\Delta_{pq}|$ is large compared to the light-atom interaction strength, we can use the Schrieffer-Wolff transformation (see Appendix \ref{sec:SW}) to eliminate the excited states. In doing so, we shall assume that $|\Delta_{pq}|$ is large compared to the hyperfine splitting in the ground and excited state manifolds and compared to the height of the optical lattice potentials. The latter ensures that the intermediate state in the Raman transition does not see the optical lattice. We then have $\Delta_{pq}\approx\Delta$ for all $p$ and $q$ and the effective Hamiltonian \eqref{HeffSW} simplifies to
\begin{equation}
H_{\rm eff}=
\frac{(Q_0VP_0)^\dag Q_0VP_0}{\hbar\Delta}
\end{equation}
to third order in $V$. We note that by choosing $\Delta$ sufficiently large, the fourth order term can be made so small that it does not contribute to the expansion to third order in $t/U$ used to derive \eqref{Heff}.

\begin{widetext}
Neglecting oscillating terms and utilizing the fact that the functions $g_p(x,y,z)$ constitute a complete set of spatial wavefunctions, we find
\begin{multline}\label{hopamp}
\langle\psi_n^{{\uparrow}_b}|H_{\rm eff}|\psi_m^{{\uparrow}_r}\rangle
=-\frac{e^2}{\hbar\Delta}
\sum_q[\langle e_q|(E_+\vec{\eta}\cdot\vec{\varepsilon}_+ +E_-\vec{\eta}\cdot\vec{\varepsilon}_-)|{\uparrow}_b\rangle]^*
E_z\langle e_q|(\vec{\eta}\cdot\vec{\varepsilon}_z)|{\uparrow}_r\rangle\\
\times\int\!\!\!\int\!\!\!\int\! f_b^*(x-x_n,y-y_n,z)\cos(k_zz)(\sin(kx)+i\sin(ky))
f_r(x-x_m,y-y_m,z) dx dy dz,
\end{multline}
\begin{multline}\label{hopblue}
\langle\psi_n^{{\uparrow}_b}|H_{\rm eff}|\psi_m^{{\uparrow}_b}\rangle=
\frac{e^2|E_z|^2}{\hbar\Delta}\sum_q|\langle e_q|(\vec{\eta}\cdot\vec{\varepsilon}_z)|{\uparrow}_b\rangle|^2
\int\!\!\!\int\!\!\!\int\! f_b^*(x-x_n,y-y_n,z)[\sin^2(kx)+\sin^2(ky)]
f_b(x-x_m,y-y_m,z) d x d y d z\\
+\frac{e^2}{\hbar\Delta}\sum_q|\langle e_q|(E_+\vec{\eta}\cdot\vec{\varepsilon}_++E_-\vec{\eta}\cdot\vec{\varepsilon}_-)|{\uparrow}_b\rangle|^2
\int\!\!\!\int\!\!\!\int\! f_b^*(x-x_n,y-y_n,z)
\cos^2(k_zz)f_b(x-x_m,y-y_m,z) d x d y d z,
\end{multline}
\begin{multline}\label{hopred}
\langle\psi_n^{{\uparrow}_r}|H_{\rm eff}|\psi_m^{{\uparrow}_r}\rangle=
\frac{e^2|E_z|^2}{\hbar\Delta}\sum_q|\langle e_q|(\vec{\eta}\cdot\vec{\varepsilon}_z)|{\uparrow}_r\rangle|^2
\int\!\!\!\int\!\!\!\int\! f_r^*(x-x_n,y-y_n,z)
[\sin^2(kx)+\sin^2(ky)]f_r(x-x_m,y-y_m,z) dx dy dz\\
+\frac{e^2}{\hbar\Delta}\sum_q|\langle e_q|(E_+\vec{\eta}\cdot\vec{\varepsilon}_+ +E_-\vec{\eta}\cdot\vec{\varepsilon}_-)|{\uparrow}_r\rangle|^2
\int\!\!\!\int\!\!\!\int\! f_r^*(x-x_n,y-y_n,z)
\cos^2(k_zz)f_r(x-x_m,y-y_m,z) dx dy dz,
\end{multline}
\end{widetext}
and the exact same set of equations with ${\uparrow}$ replaced by ${\downarrow}$. Note that only spin preserving hops are allowed due to energy conservation. These equations precisely give the hopping amplitudes $\tilde{t}_{mn}=\langle\psi_n^{\sigma_k}|H_{\rm eff}|\psi_m^{\sigma_l}\rangle$ in \eqref{Hkin}, where $\sigma_{k/l}$ is to be replaced by the relevant states. In the following two sections, we analyze first the integrals appearing in the hopping amplitudes and then the factors coming from the internal states.

\subsubsection{Spatial part of the matrix elements}

Let us first note that the integrals in \eqref{hopamp}, \eqref{hopblue}, and \eqref{hopred} decay rapidly with the distance between the sites $n$ and $m$ because the Wannier functions are localized. It is therefore sufficient to consider hops over short distances. Let us start with the integral appearing in the hopping amplitude for hops from site 1 to site 2 on the lattice (see Fig.~\ref{fig:Hkin} for the numbering of the sites), which is
\begin{multline}
-\int\!\!\!\int\!\!\!\int\! f_b^*(x-a,y,z)\cos(k_zz)(\sin(kx)+i\sin(ky))\\
\times f_r(x,y,z) dx dy dz=
-\int\!\!\!\int\!\!\!\int\! f_b^*(x-a,y,z)\cos(k_zz)\\
\times\sin(kx)f_r(x,y,z) dx dy dz
\equiv -J
\end{multline}
Note that the standing wave field in the $y$-direction does not contribute because $f_r$ and $f_b$ are even functions of $y$ whereas $\sin(ky)$ is odd. Let us compare this result to the integral
\begin{widetext}
\begin{align}
&-\int\!\!\!\int\!\!\!\int\! f_b^*(x-a,y,z)\cos(k_zz)(\sin(kx)+i\sin(ky))f_r(x-2a,y,z)d x d y d z\nonumber\\
&=-\int\!\!\!\int\!\!\!\int\! f_b^*(-x-a,y,z)\cos(k_zz)\sin(-kx)f_r(-x-2a,y,z) d x d y d z\nonumber\\
&=\int\!\!\!\int\!\!\!\int\! f_b^*(x+a,y,z)\cos(k_zz)\sin(kx)f_r(x+2a,y,z) d x d y d z\nonumber\\
&=\int\!\!\!\int\!\!\!\int\! f_b^*(x-a,y,z)\cos(k_zz)\sin(kx-2\pi)f_r(x,y,z) d x d y d z\nonumber\\
&=\int\!\!\!\int\!\!\!\int\! f_b^*(x-a,y,z)\cos(k_zz)\sin(kx)f_r(x,y,z) d x d y d z=J
\end{align}
\end{widetext}
appearing for hops from site 3 to site 2. From this we see that $\tilde{t}_{32}$ differs from $\tilde{t}_{12}$ by a minus sign as desired (see Fig.~\ref{fig:laasho}). Note that the minus sign comes from the fact that $\sin(kx)$ changes sign when displaced by one lattice constant. Similar manipulations show that the integral appearing for hops from site $L_x+2$ to site $L_x+1$ is $-J$ and that the integral appearing for hops from site $L_x+2$ to site $L_x+3$ is $J$. Since the integrals are unchanged if translated by two lattice constants in either the $x$- or the $y$-direction, it follows that the relative phases of all the nearest neighbor hops along the $x$-axis come out right.

To obtain the integrals for the hops in the $y$-direction, we only need to exchange $x$ and $y$. Since it is now $i\sin(ky)$ that contributes rather than $\sin(kx)$, we get an extra $i$ on all the integrals. The integral for a hop from site $1$ to site $L_x+1$ is thus $-i J$, for a hop from $2L_x+1$ to $L_x+1$ it is $i J$, for a hop from $L_x+2$ to $2$ it is $-i J$, and for a hop from $L_x+2$ to $2L_x+2$ it is $i J$. This is also as desired since the arrows representing the hopping amplitudes in Fig.~\ref{fig:laasho} point opposite to the hopping directions for the integrals that are $-i J$. We thus conclude that all the nearest neighbor hopping amplitudes have the correct relative phases.

Let us next consider hops between sites of the same color. Let us first note that the integral
\begin{widetext}
\begin{multline}
\int\!\!\!\int\!\!\!\int\! f_r^*(x,y,z)\sin^2(kx)f_r(x-a,y-a,z) dx dy dz
=\int\!\!\!\int\!\!\!\int\! f_r^*(x+a,y+a,z)\sin^2(kx)f_r(x,y,z) dx dy dz\\
=\int\!\!\!\int\!\!\!\int\! f_r^*(-x+a,-y+a,z)\sin^2(-kx)f_r(-x,-y,z) dx dy dz
=\int\!\!\!\int\!\!\!\int\! f_r^*(x-a,y-a,z)\sin^2(kx)f_r(x,y,z) dx dy dz
\end{multline}
\end{widetext}
is real, and the same is true if $\sin^2(kx)$ is replaced by $\sin^2(ky)$ or $\cos^2(k_zz)$, if red is replaced by blue, and if the hop is in the perpendicular direction. The hopping amplitudes for the next-nearest neighbor hops are thus all real. For the proposal to work out, we need to assume that the signs of \eqref{hopblue} and \eqref{hopred} are the same. This holds if $f_b$ and $f_r$ are Gaussian and also if the blue and the red lattices are not too different. Let us assume first that the integrals are positive. Then we would obtain the correct sign of the next-nearest neighbor hopping amplitudes for blue detuning. If the phase of \eqref{hopamp} does not come out right for the hop from site 1 to site 2, this can always be adjusted by changing the phase of either $E_z$ or of $E_+$ and $E_-$. Similar considerations apply if the integrals are negative and there is red detuning. Switching the sign of the detuning changes the sign of all the hopping amplitudes, and we note that this gives the model with reversed chirality. We can, however, get the correct chirality back simply by changing the sign of $\vec{E}_{ry}$. These considerations establish that we can get all the phases of the hopping amplitudes right, provided the phase of the internal part of \eqref{hopamp} does not depend on whether the spins are up or down. We shall see below that this is the case if $E_+$ and $E_-$ are chosen to have the same phase.

The last situation we need to consider is $n=m$. These terms give an additional contribution to the trapping. The field $\vec{E}_{bz}$ provides the trapping potential in the $z$-direction, and the other two fields modify the trapping potential in the $xy$-plane. The modification can, however, be made small by reducing $E_z$, while increasing $E_\pm$ to keep the product of the two constant.

\subsubsection{Internal part of the matrix elements}

The next important question is whether we can make the hopping rates equal for spin up and down, and for this we need to consider the internal part of the matrix elements of $H_{\rm eff}$. As for the preparation of the optical lattice potential, we shall assume that the light fields couple the ground state manifold to a $^2P_{1/2}$ orbital. The states $|e_q\rangle$ are then
\begin{align}
\left|e_1\right>
&=\sqrt{\frac{1}{3}}\left|1,-1,0,\frac{1}{2}\right\rangle_{ILS}
-\sqrt{\frac{2}{3}}\left|1,-1,1,-\frac{1}{2}\right\rangle_{ILS}, \nonumber\\
\left|e_2\right>
&=\sqrt{\frac{1}{3}}\left|1,0,0,\frac{1}{2}\right\rangle_{ILS}
-\sqrt{\frac{2}{3}}\left|1,0,1,-\frac{1}{2}\right\rangle_{ILS}, \nonumber\\
\left|e_3\right>
&=\sqrt{\frac{1}{3}}\left|1,1,0,\frac{1}{2}\right\rangle_{ILS}
-\sqrt{\frac{2}{3}}\left|1,1,1,-\frac{1}{2}\right\rangle_{ILS}, \nonumber\\
\left|e_4\right>
&=\sqrt{\frac{2}{3}}\left|1,-1,-1,\frac{1}{2}\right\rangle_{ILS}
-\sqrt{\frac{1}{3}}\left|1,-1,0,-\frac{1}{2}\right\rangle_{ILS},\nonumber\\
\left|e_5\right>
&=\sqrt{\frac{2}{3}}\left|1,0,-1,\frac{1}{2}\right\rangle_{ILS}
-\sqrt{\frac{1}{3}}\left|1,0,0,-\frac{1}{2}\right\rangle_{ILS},\nonumber\\
\left|e_6\right>
&=\sqrt{\frac{2}{3}}\left|1,1,-1,\frac{1}{2}\right\rangle_{ILS}
-\sqrt{\frac{1}{3}}\left|1,1,0,-\frac{1}{2}\right\rangle_{ILS},\nonumber
\end{align}
where the kets with subscript $ILS$ give the state in the form $|L,m_I,m_L,m_S\rangle$. The matrix elements that we shall need below are listed in table~\ref{tab:matrix}.

\begin{table}
\begin{ruledtabular}
\begin{tabular}{cccccc}
$q$ & $i$ & $|{\uparrow}_r\rangle$ & $|{\uparrow}_b\rangle$ & $|{\downarrow}_r\rangle$ & $|{\downarrow}_b\rangle$ \\
\hline
1 & $z$ & 0 & $1/3$ & $\sqrt{2}/3$ & 0 \\
2 & $z$ & $\sqrt{2}/3$ & 0 & 0 & $1/3$ \\
3 & $z$ & 0 & 0 & 0 & 0 \\
4 & $z$ & 0 & 0 & 0 & 0 \\
5 & $z$ & 0 & $-\sqrt{2}/3$ & $1/3$ & 0 \\
6 & $z$ & $-1/3$ & 0 & 0 & $\sqrt{2}/3$ \\
\hline
1 & $+$ & 0 & 0 & 0 & 0 \\
2 & $+$ & 0 & $2/3$ & $-\sqrt{2}/3$ & 0 \\
3 & $+$ & $\sqrt{2}/3$ & 0 & 0 & $-2/3$ \\
4 & $+$ & 0 & 0 & 0 & 0 \\
5 & $+$ & 0 & 0 & 0 & 0 \\
6 & $+$ & 0 & 0 & 0 & 0 \\
\hline
1 & $-$ & 0 & 0 & 0 & 0 \\
2 & $-$ & 0 & 0 & 0 & 0 \\
3 & $-$ & 0 & 0 & 0 & 0 \\
4 & $-$ & 0 & $\sqrt{2}/3$ & $2/3$ & 0 \\
5 & $-$ & $2/3$ & 0 & 0 & $\sqrt{2}/3$ \\
6 & $-$ & 0 & 0 & 0 & 0
\end{tabular}
\end{ruledtabular}
\caption{$\langle e_q|\vec{\eta}\cdot\vec{\varepsilon}_i|\psi\rangle$ in units of $R/\sqrt{3}$, where $|\psi\rangle\in\{|{\uparrow}_r\rangle,|{\uparrow}_b\rangle, |{\downarrow}_r\rangle,|{\downarrow}_b\rangle\}$ and $R$ is the radial integral in \eqref{R}.}\label{tab:matrix}
\end{table}

The internal part of the matrix element describing nearest neighbor hops of spin up fermions is
\begin{multline}
\sum_q[\langle e_q|(E_+\vec{\eta}\cdot\vec{\varepsilon}_+ +E_-\vec{\eta}\cdot\vec{\varepsilon}_-)|{\uparrow}_b\rangle]^*
E_z\langle e_q|(\vec{\eta}\cdot\vec{\varepsilon}_z)|{\uparrow}_r\rangle=\\
\frac{2\sqrt{2}|R|^2}{27}E_+^*E_z
\end{multline}
whereas it is
\begin{multline}
\sum_q[\langle e_q|(E_+\vec{\eta}\cdot\vec{\varepsilon}_+ +E_-\vec{\eta}\cdot\vec{\varepsilon}_-)|{\downarrow}_b\rangle]^*
E_z\langle e_q|(\vec{\eta}\cdot\vec{\varepsilon}_z)|{\downarrow}_r\rangle=\\
\frac{\sqrt{2}|R|^2}{27}E_-^*E_z
\end{multline}
for spin down. This shows that the hopping amplitudes are the same for spin up and spin down fermions provided we choose $E_-=2E_+$. We assume this to be the case in the following.

Let us next consider hops between blue lattice sites. Note that
\begin{equation}
\sum_q|\langle e_q|(\vec{\eta}\cdot\vec{\varepsilon}_z)|{\uparrow}_b\rangle|^2
=\sum_q|\langle e_q|(\vec{\eta}\cdot\vec{\varepsilon}_z)|{\downarrow}_b\rangle|^2=\frac{1}{9}|R|^2
\end{equation}
and
\begin{multline}
\sum_q|\langle e_q|(E_+\vec{\eta}\cdot\vec{\varepsilon}_+
+E_-\vec{\eta}\cdot\vec{\varepsilon}_-)|{\uparrow}_b\rangle|^2=\\
\sum_q|\langle e_q|(E_+\vec{\eta}\cdot\vec{\varepsilon}_+
+E_-\vec{\eta}\cdot\vec{\varepsilon}_-)|{\downarrow}_b\rangle|^2
=\frac{4}{9}|R|^2|E_+|^2.
\end{multline}
The next-nearest neighbor hopping thus happens at the same rate for fermions in the blue spin up state as for fermions in the blue spin down state.

For hops between red sites we get
\begin{equation}
\sum_q|\langle e_q|(\vec{\eta}\cdot\vec{\varepsilon}_z)|{\uparrow}_r\rangle|^2
=\sum_q|\langle e_q|(\vec{\eta}\cdot\vec{\varepsilon}_z)|{\downarrow}_r\rangle|^2=\frac{1}{9}|R|^2
\end{equation}
and
\begin{multline}
\sum_q|\langle e_q|(E_+\vec{\eta}\cdot\vec{\varepsilon}_+
+E_-\vec{\eta}\cdot\vec{\varepsilon}_-)|{\uparrow}_r\rangle|^2=\\
\sum_q|\langle e_q|(E_+\vec{\eta}\cdot\vec{\varepsilon}_+
+E_-\vec{\eta}\cdot\vec{\varepsilon}_-)|{\downarrow}_r\rangle|^2
=\frac{2}{3}|R|^2|E_+|^2,
\end{multline}
so that the hopping rate is again the same for spin up and spin down. If the red and blue optical lattices were the same, the above results also show that hops would occur at a faster rate on the red lattice than on the blue lattice. This can be avoided, however, by making the difference between the maximum and the minimum value of the red lattice potential larger than the same difference for the blue lattice, i.e., by choosing $\alpha^2$ slightly larger than $1$ in Fig.~\ref{fig:potential}.

\subsection{Implementation of the interaction terms}

Finally, we need to account for the on-site interaction terms. As in the standard Fermi-Hubbard model, interactions between atoms on the same site occur naturally, and Feshbach resonances can be used to tune the interaction strength over a wide range of values \cite{kohl,jordens,schneider,giorgini,buchler}. Let us also note here that the intensity of the laser beams used to create the optical lattice influences the relative strength of the interaction terms and the tunneling terms \cite{jaksch98}. Increasing the intensity, increases the potential barrier between sites and therefore reduces the tunneling rate. At the same time, a larger intensity also reduces the spatial width of the Wannier functions and this increases the interaction strength between two atoms on the same site. Finally, the rate of laser assisted tunneling can be adjusted independently by varying the amplitudes of the lasers driving the Raman transitions.

\section{Conclusion}\label{sec:conclusion}

In conclusion, we have described a scheme to implement a bosonic FQH-like state in ultracold fermions in optical lattices. The FQH-like state appears as the ground state of a Fermi-Hubbard-like Hamiltonian with complex nearest neighbor and real next-nearest neighbor hopping terms for suitable parameters in the Mott insulating regime. The proposal uses a checkerboard optical lattice and laser assisted tunneling with a suitable configuration of laser beams. The experimental requirements are similar to those needed to observe the N\'eel antiferromagnetic ordering in the standard Fermi-Hubbard Mott insulator combined with the implementation of laser assisted tunneling in this system. The model can thus be implemented with present or planned technology.

\begin{acknowledgments}
This work has been supported by the EU project SIQS, FIS2012-33642, QUITEMAD (CAM), and the Severo Ochoa Program.
\end{acknowledgments}

\appendix

\section{The Schrieffer-Wolff transformation}\label{sec:SW}

A detailed description of the Schrieffer-Wolff transformation can be found in \cite{bravyi}, and here we give only a brief overview, summarizing the results needed in this paper. Consider a quantum system with Hamiltonian $H_0$ and let $|i\rangle$ denote the eigenstates of $H_0$ with eigenvalues $E_i$, i.e., $H_0|i\rangle=E_i|i\rangle$. Let $\mathcal{P}_0$ be the subspace spanned by all states $|i\rangle$ for which $E_i$ belongs to a given energy interval, and let $\mathcal{Q}_0$ be the subspace spanned by all other states. We shall assume that the states in $\mathcal{P}_0$ are separated by an energy gap from the states in $\mathcal{Q}_0$, i.e., $|E_i-E_j|\geq\hbar\Delta_0$ for all $|i\rangle\in\mathcal{P}_0$ and all $|j\rangle\in\mathcal{Q}_0$, where $\hbar\Delta_0$ is some constant larger than zero. Typically, the interval is chosen to encompass the lowest energy states, and for convenience we shall therefore refer to $\mathcal{P}_0$ as the low energy subspace even though the results are more general.

We now add a perturbation $V$, such that the Hamiltonian is $H=H_0+V$, and define $\mathcal{P}$ ($\mathcal{Q}$) to be the low (high) energy subspace with respect to $H$. If $V$ is small enough that it changes all of the energies $E_i$ in the spectrum of $H_0$ by less than $\hbar\Delta_0/2$, we can choose the dimension of $\mathcal{P}$ to be the same as the dimension of $\mathcal{P}_0$, and there will be an energy gap between the low and the high energy states of $H$. In this case, one can find a unitary transformation $U$ that transforms $\mathcal{P}$ into $\mathcal{P}_0$ and $\mathcal{Q}$ into $\mathcal{Q}_0$. Applying this transformation to $H$ is the Schrieffer-Wolff transformation, and the resulting Hamiltonian is block diagonal with respect to $\mathcal{P}_0$ and $\mathcal{Q}_0$. One can then obtain an effective Hamiltonian $H_{\rm eff}$ for the low energy physics by discarding the block of the Hamiltonian acting on $\mathcal{Q}_0$.

When the perturbation is small, $U$ is close to the identity, and one can write $U=e^S$ and Taylor expand $S$ in $V$. As derived in \cite{bravyi}, this leads to
\begin{equation}\label{HeffSW}
H_{\rm eff}=\sum_{n}H_{\rm eff}^{(n)},
\end{equation}
where $H_{\rm eff}^{(n)}$ is $n$th order in $V$ and $H_{\rm eff}^{(n)}$ for $n\leq3$ is given by
\begin{align}\label{Heffn}
H_{\rm eff}^{(0)}&=P_0H_0P_0,\\
H_{\rm eff}^{(1)}&=P_0VP_0,\\
H_{\rm eff}^{(2)}&=\frac{1}{2}\sum_{i,j}\frac{P_0|i\rangle\langle i|P_0VQ_0|j\rangle\langle j|Q_0VP_0}{E_i-E_j}+h.c.\\
H_{\rm eff}^{(3)}&=\frac{1}{2}
\sum_{i,j,k}\frac{P_0VQ_0|j\rangle\langle j|Q_0VQ_0|i\rangle\langle i|Q_0VP_0|k\rangle\langle k|P_0}{(E_i-E_k)(E_j-E_k)}\nonumber\\
&\hspace{-6mm}-\frac{1}{2}
\sum_{i,j,k}\frac{P_0VQ_0|k\rangle\langle k|Q_0VP_0|i\rangle\langle i|P_0VP_0|j\rangle\langle j|P_0}{(E_i-E_k)(E_j-E_k)}+h.c.\label{Heff3}
\end{align}
Here, $P_0$ ($Q_0$) is the projector onto $\mathcal{P}_0$ ($\mathcal{Q}_0$), and $h.c.$ is the hermitian conjugate.

\section{Effective model in the Mott insulating regime}\label{sec:effmodel}

We derive the effective Hamiltonian of the Fermi-Hubbard-like model in \eqref{FH} in the limit $|t|,|t'|\ll U$ by applying the Schrieffer-Wolff transformation in Appendix \ref{sec:SW}. In this case, $H_0=H_{\rm int}$ and $V=\sum_{\sigma}H_{{\rm kin},\sigma}$. The low energy subspace $\mathcal{P}_0$ of $H_0$ consists of all states with precisely one fermion on each site, and these states have zero energy. The lowest excited states have precisely one site with double occupation and have energy $U$. In the following, we compute the terms \eqref{Heffn} in the effective Hamiltonian up to third order in $V$.

\subsection{Zeroth and first order term}

Since all the low energy states have precisely one fermion at each site, we immediately get
\begin{equation}
H_{\rm eff}^{(0)}=H_{\rm eff}^{(1)}=0.
\end{equation}

\subsection{Second order term}

The second order term describes processes in which we start from a state in $\mathcal{P}_0$ and apply the potential twice, after which we must be back to a state in $\mathcal{P}_0$. The only possibility is therefore a fermion hopping from site $n$ to site $m$ followed by a fermion hopping from $m$ to $n$. We can therefore treat the terms in $V$ corresponding to hops between different pairs of lattice sites independently. Considering one of these terms, the relevant part of the potential is
\begin{equation}
V_{mn}\equiv\sum_{\sigma}(\tilde{t}_{mn}a_{n\sigma}^\dag a_{m\sigma} +\tilde{t}^*_{mn}a_{m\sigma}^\dag a_{n\sigma}),
\end{equation}
where $n$ and $m$ must be nearest or next-nearest neighbors on the lattice. $V_{mn}$ gives the contribution
\begin{multline}
H_{{\rm eff},mn}^{(2)}\equiv
-\frac{|\tilde{t}_{mn}|^2}{U}\sum_{\sigma}\sum_{\sigma'} P_0(a_{m,\sigma}^\dag a_{n,\sigma}a_{n,\sigma'}^\dag a_{m,\sigma'}\\
+a_{n,\sigma}^\dag a_{m,\sigma}a_{m,\sigma'}^\dag a_{n,\sigma'})P_0
\end{multline}
to $H_{\rm eff}^{(2)}$.

We would like to express $H_{{\rm eff},mn}^{(2)}$ in terms of spin operators, and we therefore investigate the action of $H_{{\rm eff},mn}^{(2)}$ on the states $|{\uparrow}_n{\uparrow}_m\rangle$, $|{\uparrow}_n{\downarrow}_m\rangle$, $|{\downarrow}_n{\uparrow}_m\rangle$, and $|{\downarrow}_n{\downarrow}_m\rangle$ in $\mathcal{P}_0$, where $|{\uparrow}_n{\downarrow}_m\rangle$, e.g., represents the state with the fermion on site $n$ in the ${\uparrow}$ state and the fermion on site $m$ in the ${\downarrow}$ state, whereas the state of all the other spins is not important and therefore not specified. Using the anticommutation relations of the fermion operators, we get
\begin{align}\label{heffac}
H_{{\rm eff},mn}^{(2)}|{\uparrow}_n{\uparrow}_m\rangle&=0\\
H_{{\rm eff},mn}^{(2)}|{\uparrow}_n{\downarrow}_m\rangle &=-\frac{2|\tilde{t}_{mn}|^2}{U}(|{\uparrow}_n{\downarrow}_m\rangle-|{\downarrow}_n{\uparrow}_m\rangle) \nonumber\\
H_{{\rm eff},mn}^{(2)}|{\downarrow}_n{\uparrow}_m\rangle
&=-\frac{2|\tilde{t}_{mn}|^2}{U}(|{\downarrow}_n{\uparrow}_m\rangle-|{\uparrow}_n{\downarrow}_m\rangle) \nonumber\\
H_{{\rm eff},mn}^{(2)}|{\downarrow}_n{\downarrow}_m\rangle&=0.\nonumber
\end{align}
Since $H_{{\rm eff},mn}^{(2)}$ is $SU(2)$ invariant, one would expect that it can be written in terms of $\vec{S}_n\cdot\vec{S}_m$. A small computation shows that
\begin{equation}
H_{{\rm eff},mn}^{(2)}=\frac{2|\tilde{t}_{mn}|^2}{U}
\left(2\vec{S}_n\cdot\vec{S}_m-\frac{1}{2}\right)
\end{equation}
indeed reproduces \eqref{heffac}. To get $H_{\rm eff}^{(2)}$ in the spin basis we only need to sum over all pairs of nearest and next-nearest neighbors.

\subsection{Third order term}

Since $P_0VP_0=0$, the second term on the right hand side of \eqref{Heff3} vanishes. All nonzero contributions to $H_{\rm eff}^{(3)}$ then involve hops between three sites, where both of the intermediate states have energy $U$, and therefore \eqref{Heff3} simplifies to
\begin{equation}
H_{\rm eff}^{(3)}=\frac{1}{U^2}P_0VQ_0VQ_0VP_0.
\end{equation}
The three sites must pairwise be nearest or next-nearest neighbors. Let us consider a triangle with vertices labeled $n$, $m$, $p$ when going around the triangle in the counter clockwise direction. The contribution to $H_{\rm eff}^{(3)}$ from this triangle is
\begin{multline}
 H_{{\rm eff},nmp}^{(3)}\equiv
\frac{i t^2t'}{U^2}\sum_{\sigma,\sigma',\sigma''} \sum_{\substack{{\rm All\ permuta-}\\{\rm tions\ of\ }n,m,p}}
P_0a_{n,\sigma}^\dag a_{h(n),\sigma}\\
\times a_{m,\sigma'}^\dag a_{h(m),\sigma'}
a_{p,\sigma''}^\dag a_{h(p),\sigma''}P_0\\
-\frac{i t^2t'}{U^2}\sum_{\sigma,\sigma',\sigma''} \sum_{\substack{\rm All\ permuta-\\{\rm tions\ of\ }n,m,p}}
P_0a_{n,\sigma}^\dag a_{h^{-1}(n),\sigma}\\
\times a_{m,\sigma'}^\dag a_{h^{-1}(m),\sigma'}
a_{p,\sigma''}^\dag a_{h^{-1}(p),\sigma''}P_0,
\end{multline}
where the function $h$ is defined such that $h(n)=m$, $h(m)=p$, and $h(p)=n$ and $h^{-1}$ is the inverse of $h$. The action of $H_{{\rm eff},nmp}^{(3)}$ on the spin states in $\mathcal{P}_0$ is
\begin{align}
H_{{\rm eff},nmp}^{(3)}|{\uparrow}_n{\uparrow}_m{\uparrow}_p\rangle&=0,\\
H_{{\rm eff},nmp}^{(3)}|{\uparrow}_n{\uparrow}_m{\downarrow}_p\rangle
&=-\frac{6i t^2t'}{U^2}|{\uparrow}_n{\downarrow}_m{\uparrow}_p\rangle
+\frac{6i t^2t'}{U^2}|{\downarrow}_n{\uparrow}_m{\uparrow}_p\rangle,\nonumber\\
H_{{\rm eff},nmp}^{(3)}|{\uparrow}_n{\downarrow}_m{\uparrow}_p\rangle
&=-\frac{6i t^2t'}{U^2}|{\downarrow}_n{\uparrow}_m{\uparrow}_p\rangle
+\frac{6i t^2t'}{U^2}|{\uparrow}_n{\uparrow}_m{\downarrow}_p\rangle,\nonumber\\
H_{{\rm eff},nmp}^{(3)}|{\downarrow}_n{\uparrow}_m{\uparrow}_p\rangle
&=-\frac{6i t^2t'}{U^2}|{\uparrow}_n{\uparrow}_m{\downarrow}_p\rangle
+\frac{6i t^2t'}{U^2}|{\uparrow}_n{\downarrow}_m{\uparrow}_p\rangle,\nonumber\\
H_{{\rm eff},nmp}^{(3)}|{\downarrow}_n{\downarrow}_m{\uparrow}_p\rangle
&=-\frac{6i t^2t'}{U^2}|{\downarrow}_n{\uparrow}_m{\downarrow}_p\rangle
+\frac{6i t^2t'}{U^2}|{\uparrow}_n{\downarrow}_m{\downarrow}_p\rangle,\nonumber\\
H_{{\rm eff},nmp}^{(3)}|{\downarrow}_n{\uparrow}_m{\downarrow}_p\rangle
&=-\frac{6i t^2t'}{U^2}|{\uparrow}_n{\downarrow}_m{\downarrow}_p\rangle
+\frac{6i t^2t'}{U^2}|{\downarrow}_n{\downarrow}_m{\uparrow}_p\rangle,\nonumber\\
H_{{\rm eff},nmp}^{(3)}|{\uparrow}_n{\downarrow}_m{\downarrow}_p\rangle
&=-\frac{6i t^2t'}{U^2}|{\downarrow}_n{\downarrow}_m{\uparrow}_p\rangle
+\frac{6i t^2t'}{U^2}|{\downarrow}_n{\uparrow}_m{\downarrow}_p\rangle,\nonumber\\
H_{{\rm eff},nmp}^{(3)}|{\downarrow}_n{\downarrow}_m{\downarrow}_p\rangle&=0,\nonumber
\end{align}
where again we only specify the states of the spins on which $H_{{\rm eff},nmp}^{(3)}$ acts. Given the $SU(2)$ symmetry, we compare these results to the action of $\vec{S}_n\cdot\left(\vec{S}_m \times \vec{S}_p\right)$ and get
\begin{equation}
H_{{\rm eff},nmp}^{(3)}=-\frac{24t^2t'}{U^2}
\vec{S}_n\cdot\left(\vec{S}_m \times \vec{S}_p\right).
\end{equation}
The complete third order term $H_{\rm eff}^{(3)}$ is then obtained by summing $H_{{\rm eff},nmp}^{(3)}$ over all triangles.

\bibliography{bibfil}

\end{document}